\documentclass[twocolumn,aps,pra,superscriptaddress,showpacs]{revtex4-1}
\usepackage{epsfig}
\usepackage[colorlinks=true,citecolor=blue,urlcolor=blue,linkcolor=blue]{hyperref}
\usepackage{chemscheme}
\usepackage{graphicx}
\usepackage{multirow}
\usepackage[table,xcdraw]{xcolor}
\usepackage{natbib}
\usepackage{chemfig}
\usepackage[version=4]{mhchem}
\usepackage[normalem]{ulem}
\usepackage{floatrow}
\usepackage{tabulary}
\usepackage{booktabs}
\usepackage[flushleft]{threeparttable}
\usepackage{multirow}
\usepackage{xspace}
\usepackage{xcolor} 
\definecolor{darkgreen}{rgb}{0.0, 0.6, 0.0}

\DeclareMathAlphabet{\mathbsf}{OT1}{cmss}{bx}{n}

\newcommand{\phenyl}{\raisebox{0.21ex}{-}\raisebox{0.75ex}{\scalebox{0.175}{\chemfig[height=1.7ex]{[:-30]**6(------)}}}}
\newcommand{\phenylX}[1]{\raisebox{0.21ex}{-}\raisebox{0.75ex}{\scalebox{0.175}{\chemfig[height=1.7ex]{[:-30]**6(------)}}}\raisebox{0.21ex}{-}#1}
\newcommand{\cm}{cm$^{-1}$\xspace}
\newcommand{\A}{$\widetilde A$\xspace}

\newcommand{\X}{$\widetilde X$\xspace}

\begin{document}

\title{Functionalizing Aromatic Compounds with Optical Cycling Centers}
\author{Guo-Zhu Zhu}
\affiliation{Department of Physics and Astronomy, University of California, Los Angeles, California 90095, USA}

\author{Debayan Mitra}
\altaffiliation{Present address: Department of Physics, Columbia University, New York}
\affiliation{Harvard-MIT Center for Ultracold Atoms, Cambridge, MA 02138, USA}
\affiliation{Department of Physics, Harvard University, Cambridge, MA 02138, USA}

\author{Benjamin L. Augenbraun}
\affiliation{Harvard-MIT Center for Ultracold Atoms, Cambridge, MA 02138, USA}
\affiliation{Department of Physics, Harvard University, Cambridge, MA 02138, USA}

\author{Claire E. Dickerson}
\affiliation{Department of Chemistry and Biochemistry, University of California, Los Angeles, California 90095, USA}

\author{Michael J. Frim}
\affiliation{Department of Physics, Harvard University, Cambridge, MA 02138, USA}

\author{Guanming Lao}
\affiliation{Department of Physics and Astronomy, University of California, Los Angeles, California 90095, USA}

\author{Zack D. Lasner}
\affiliation{Harvard-MIT Center for Ultracold Atoms, Cambridge, MA 02138, USA}
\affiliation{Department of Physics, Harvard University, Cambridge, MA 02138, USA}

\author{Anastassia N. Alexandrova}
\affiliation{Department of Chemistry and Biochemistry, University of California, Los Angeles, California 90095, USA}
\affiliation{Center for Quantum Science and Engineering, University of California, Los Angeles, California 90095, USA}

\author{Wesley C. Campbell}
\affiliation{Department of Physics and Astronomy, University of California, Los Angeles, California 90095, USA}
\affiliation{Center for Quantum Science and Engineering, University of California, Los Angeles, California 90095, USA}
\affiliation{Challenge Institute for Quantum Computation, University of California, Los Angeles, California 90095, USA}

\author{Justin R. Caram}
\affiliation{Department of Chemistry and Biochemistry, University of California, Los Angeles, California 90095, USA}
\affiliation{Center for Quantum Science and Engineering, University of California, Los Angeles, California 90095, USA}

\author{John M. Doyle}
\affiliation{Harvard-MIT Center for Ultracold Atoms, Cambridge, MA 02138, USA}
\affiliation{Department of Physics, Harvard University, Cambridge, MA 02138, USA}

\author{Eric R. Hudson}
\affiliation{Department of Physics and Astronomy, University of California, Los Angeles, California 90095, USA}
\affiliation{Center for Quantum Science and Engineering, University of California, Los Angeles, California 90095, USA}
\affiliation{Challenge Institute for Quantum Computation, University of California, Los Angeles, California 90095, USA}

\date{\today}

\begin{abstract}
Molecular design principles provide guidelines for augmenting a molecule with a smaller group of atoms to realize a desired property or function. We demonstrate that these concepts can be used to create an optical cycling center that can be attached to a number of aromatic ligands, allowing the scattering of many photons from the resulting molecules without changing the molecular vibrational states. We provide further design principles that indicate the ability to expand this work. This represents a significant step towards a quantum functional group, which may serve as a generic qubit moiety that can be attached to a wide range of molecular structures and surfaces.

\end{abstract}

\maketitle

Molecules and surfaces can be augmented 
with molecular fragments that imbue a desired property to the system. 
Such functionalization can determine the system reactivity and properties, allowing a host of capabilities, such as catalysis~\cite{campisciano2019modified, wang2019catalysis} and biological and chemical binding and sensing~\cite{liu2012biological,saha2012goldsensor,guillemard2021late}.
As the properties of the functional groups are often not strongly affected by the bonding to the host molecule, the technique can bring the same function to a wide-variety of systems.

It is interesting to consider extending the idea of a functional group to new operations. 
For example, can a robust qubit moiety be designed to act as a functional group attached to a larger molecule? 
Can multiple such \emph{quantum functional groups} connect through space, allowing the host compound to serve as a bus for entanglement? 
And, relatedly, can a quantum functional group be used to control or witness the dynamics of a larger molecular whole?

Here, we shed light on these questions by demonstrating the key features of a quantum functional group, known as an optical cycling center (OCC). 
We experimentally show that a functional group, an alkaline-earth(I)-oxygen moiety, can be attached to a variety of aromatic compounds while retaining the property that it can scatter many photons without changing vibrational state. 
This property opens the door to using lasers to cool~\cite{di2004laser} external and internal degrees of freedom of large molecules by simply functionalizing them with an OCC. 
With this capability come the prerequisites for quantum information science with large molecules, namely mechanical control and qubit state preparation and readout. 

The process for OCC functionalization can be understood intuitively by considering the neutral molecules that have previously been laser cooled. 
Much success has been had with diatomic molecules composed of an alkaline-earth metal atom bonded to a halogen atom~\cite{shuman2010laser,anderegg2017radio}. 
In such molecules, the halogen atom withdraws an electron from the alkaline-earth atom, leaving one metal-centered radical electron whose highest-occupied and lowest-unoccupied molecular orbitals, HOMO and LUMO, respectively, do not strongly participate in the molecular bonding and closely resemble those of an alkali atom. 
The result is a molecule that can absorb and emit many photons without changing vibrational state~\cite{isaev2016polyatomic}. 

More recently, this theme was demonstrated experimentally with an alkaline-earth metal atom bonded to -\chemfig{OH}~\cite{vilas2021CaOHMOT}.
It was proposed that the same approach could be used for an entire class of alkaline metal-oxide-radical molecules, including complex polyatomic molecules~\cite{Kozyryev2016prop}, and later laser cooling was extended to the -\chemfig{OCH_3} ligand~\cite{mitra2020direct}.
Like a halogen atom, all of these fragments withdraw an electron from the metal atom. 
This suggests that the alkaline-earth(I)-oxygen moiety can be considered as a functional group, akin to an alcohol or ether.
Subsequent bonding of this functional group to electron withdrawing ligands should therefore allow scattering of many photons without changing the vibrational state of the resulting molecule~\cite{dickerson2021franck,dickerson2021optical}. 

\begin{figure*}
    \centering
    \includegraphics[width = 0.8\textwidth]{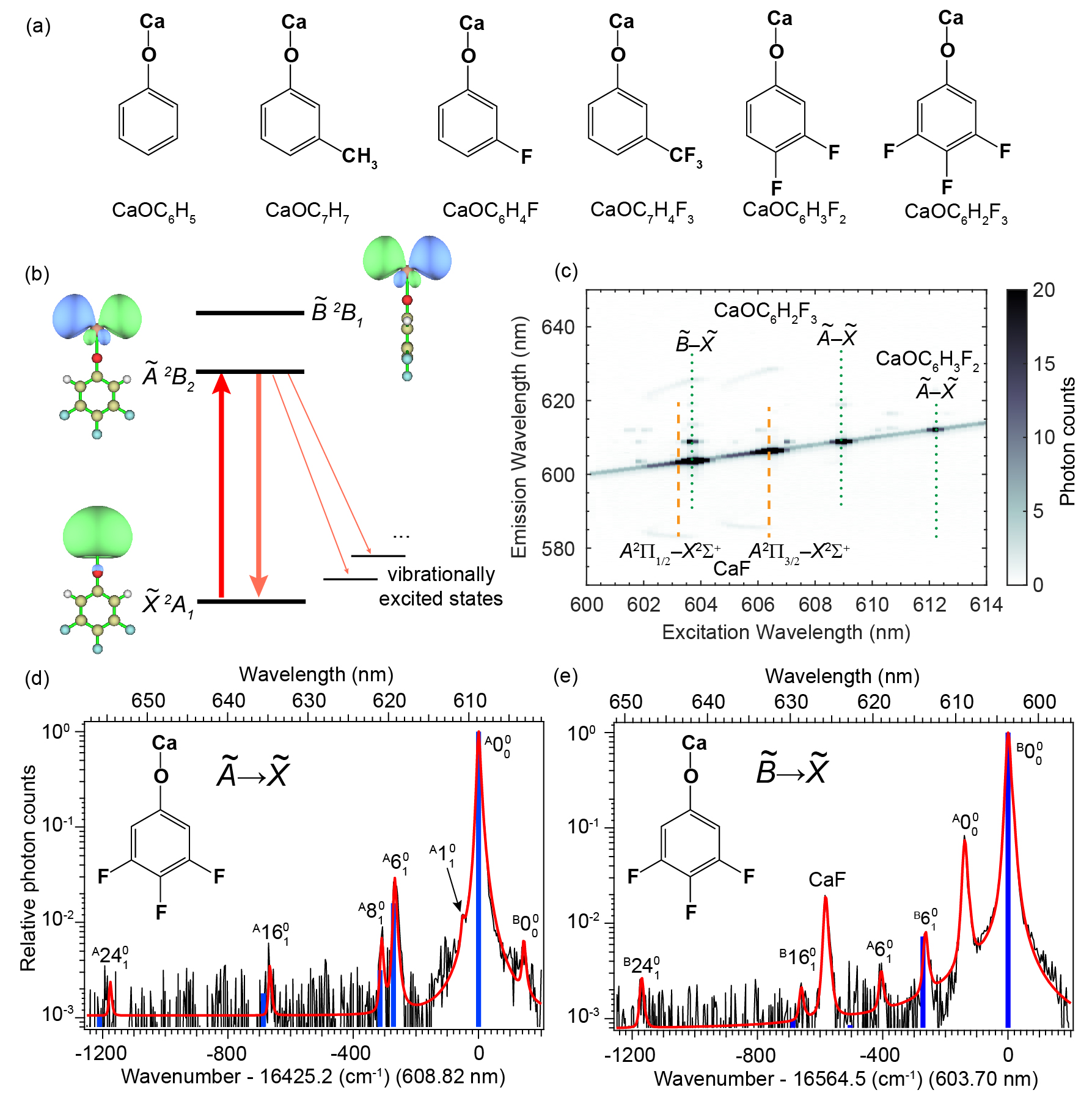}
    \caption{ (a) Molecular structures of calcium phenoxide and derivatives. (b) Molecular orbital and schematic energy levels of CaO\protect\phenyl. All other molecules have similar orbitals and energy levels. (c) 2D DLIF spectrum following the reaction of Ca with 3,4,5-trifluorophenol. The orange dashed lines indicate the bands of CaF while green dotted lines indicate the bands of CaO\protect\phenylX{345F} and CaO\protect\phenylX{34F}.
    DLIF spectra of CaO\protect\phenylX{345F} when exciting the (d) $\widetilde A \leftarrow \widetilde X$ at 608.82 nm and (e) $\widetilde B \leftarrow \widetilde X$ at 603.70 nm. Experimental data (black) is overlaid with a Pearson distribution fit (red). The blue vertical lines in the DLIF spectra indicate the calculated frequencies of the vibrational modes and the height of the lines reflect their calculated relative strengths.
    }
    \label{fig:Structure_2D}
\end{figure*}

To explore the limits of this concept, we studied the functionalization of aromatic compounds with an OCC.
Specifically, we attached a Ca(I)$-$O unit to a phenyl group (\phenyl) and its derivatives \phenylX{X} (X= $m$CH$_3$, $m$F, $m$CF$_3$, 34F and 345F, Fig.~\ref{fig:Structure_2D}(a)), and measured the vibronic spectra of the resulting molecules with dispersed laser-induced fluorescence (DLIF) spectroscopy.
We found that the OCC transition frequency is linearly related to the acid dissociation constant (pK$_a$) of the precursor compound, providing a simple means for predicting molecular properties.
Further, the vibrational branching ratios (VBRs) were determined and regardless of the choice of ligand it was found that $\gtrsim90\%$ of the photon scattering events did not change the vibrational state of the molecule.
Small variation in the VBR, at the few percent level, was observed, and found to be consistent with a theoretical calculation suggesting that pK$_a$ provides a simple guide for designing ligands with the best OCC performance \cite{dickerson2021franck}. 
From this work, we found that CaO\phenylX{345F} does not change its vibrational state during roughly 99\% of scattering events, implying that laser cooling is possible with current technology.
In what follows, we describe the apparatus and experiments, present the recorded spectra and measured VBRs, and discuss the next steps for decorating molecules and surfaces with quantum functional groups.  

\begin{figure*}[ht!]
    \centering
    \includegraphics[scale=1]{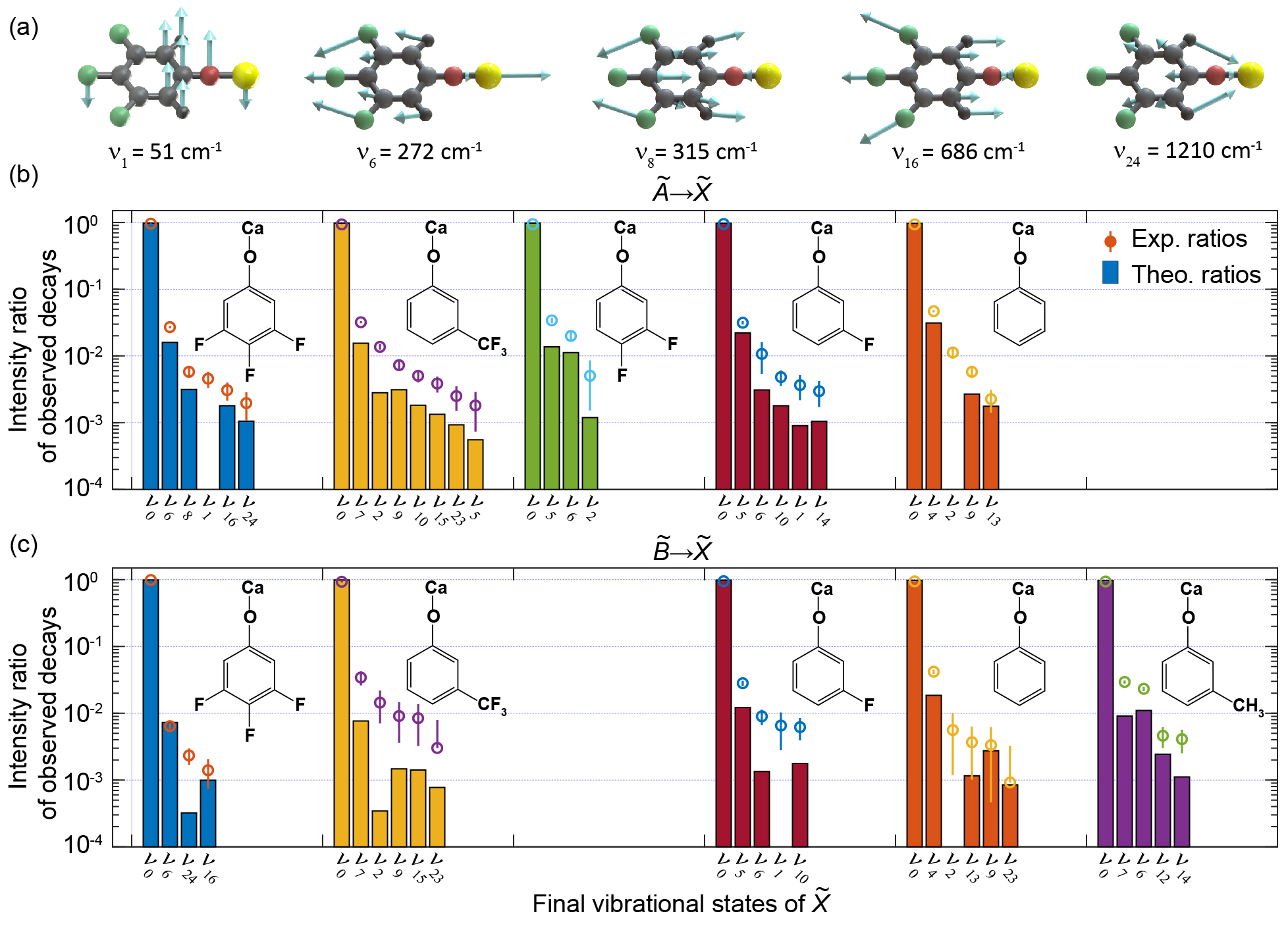}
    \caption{(a) The observed vibrational modes of CaO\protect\phenylX{345F}. Arrows indicate the direction of vibrational displacements. Intensity ratio of observed decays, relative to total observed decays, for (b) $\widetilde{A}\rightarrow \widetilde{X}$ and (c) $\widetilde{B}\rightarrow \widetilde{X}$ transitions for all molecules and modes studied in this work, arranged in order of increasing pK$_a$. 
    Experimental values are denoted with circles while calculated values are depicted as bars for clarity. Error bars are statistical standard errors. The vibrational mode denoted as $\nu_I$ indicates the decay to the final vibrational state of $I^0_1$.
    The $\widetilde{A}\rightarrow \widetilde{X}$ decay of CaO\protect\phenylX{$m$CH$_3$} and $\widetilde{B}\rightarrow \widetilde{X}$ decay of CaO\protect\phenylX{34F} are omitted due to coincidences with CaOH and CaF decays, respectively.
    }
    \label{fig.data_plus_theory}
\end{figure*}

The molecules were produced by the reaction of Ca atoms with ligand precursors and cooled in a cryogenic He buffer-gas cell operated at $\approx$~9~K~\cite{hutzler2012buffer} (see SI for details).
A tunable, pulsed optical parametric oscillator (OPO) provided the excitation light and a monochromator augmented with an intensified charge-coupled device camera (ICCD) recorded the dispersed fluorescence (see SI for details). 

Molecular species were first identified via 2D spectroscopy, performed by scanning the OPO frequency and recording the DLIF. 
Transitions between the ground ($\widetilde X\,{}^2\!A_1$) and two lowest-energy electronic states ($\widetilde A\, {}^2\!B_2$ and $\widetilde B\,{}^2\!B_1$) have been predicted to have potential for laser cooling~\cite{dickerson2021franck} due to the localization of the molecular orbitals on the Ca atom (see Fig. \ref{fig:Structure_2D}(b)), and are therefore the targets of this work.
Example data is shown in Fig.~\ref{fig:Structure_2D}(c) for CaO\phenylX{345F}, while the 2D spectra of all other molecules observed are shown in the SI.
Four strong spectral features were observed. 
The excitation to the $\widetilde{B}$ electronic state of CaO\phenylX{345F} is indicated by the green dotted line near 603.5~nm, while the excitation to the $\widetilde{A}$ electronic state is at 608.5~nm.
The two broad bands near 603.5 nm and 606.5 nm, marked by dashed orange lines, are attributed to the $v'=0 - v''=0$ bands of the $A\,^2\Pi_{1/2} \leftarrow X\,^2\Sigma^+$ and $A\,^2\Pi_{3/2} \leftarrow X\,^2\Sigma^+$ transitions, respectively, of CaF, which is also formed when Ca reacts with a fluorinated phenol.
A weaker feature at 612.0~nm is due to the $\widetilde A - \widetilde X$ transition of CaO\phenylX{34F}, which is a by-product of the reaction between Ca and 3,4,5-trifluorophenol. 
Assignments were made by comparing observed and calculated (see SI) vibrational frequencies, as discussed below.

Having identified the molecules, DLIF measurements were recorded by tuning the OPO to a selected resonance and accumulating between 4000-8000 ICCD exposures.
Representative DLIF spectra for CaO\phenylX{345F} are shown in Figs.~\ref{fig:Structure_2D}(d-e), while those of the other molecules are presented in the SI.
All spectra are plotted in terms of the energy difference (in \cm) relative to the excitation energy and are normalized to the peak at the origin.
Figure~\ref{fig:Structure_2D}(d) shows the DLIF spectrum of CaO\phenylX{345F} when exciting the $\widetilde A \leftarrow \widetilde X$ transition at 608.82 nm. 
The peak, labeled as $^A0^0_0$, represents the decay from the excited $\widetilde{A}(v'=0)$ state to the ground $\widetilde{X}(v''=0)$ state. 
The strongest vibration-changing decay, observed at $-267$ \cm\ and labeled $^A6^0_1$, is assigned to the Ca$-$O stretching mode of $\widetilde{X}$ state with a predicted harmonic frequency of 272 \cm (Fig.~\ref{fig.data_plus_theory}(a)). 
Similarly, peaks at $-307$ \cm and $-670$ \cm can be assigned to $8^0_1$ and $16^0_1$, respectively, which are both symmetric stretching modes involving the benzene ring (Fig.~ \ref{fig.data_plus_theory}(a)). 
A weak decay at $-1184$ \cm is attributed to the high-frequency stretching mode $24^0_1$ (Fig.~\ref{fig.data_plus_theory}(a)).
The small shoulder next to the diagonal peak is due to decays to the lowest-frequency fundamental bending mode, $1^0_1$. 
The vertical blue lines are the calculated VBRs normalized by the predicted value for the 0-0 decay.
Interestingly, as noted by the absence of a predicted VBR for the $1^0_1$ peak, theoretical calculations predict this decay pathway to be negligible.
The observed strength of this decay is likely due to vibronic couplings among and anharmonicities within the low-frequency modes \cite{domcke1977theoFCF, Fischer1984, dickerson2021franck} not considered in the present calculation.

The DLIF spectrum of CaO\phenylX{345F} from the $\widetilde{B}$~state is shown in Fig. \ref{fig:Structure_2D}(e). 
In addition to the non-vibration-changing decay $^B0^0_0$ and the dominant vibration-changing decay $^B6^0_1$, decays are observed with shifts of $-139$ \cm and $-407$ \cm.
These peaks are due to emissions from $\widetilde{A}(v'=0)$, which is presumably populated by collisional relaxation in the buffer gas~\cite{Liu2019collision,dagdigian1997collision}. 
Interestingly, we also see evidence of collision-induced excitation when exciting to the \A\ state; as seen in Fig. \ref{fig:Structure_2D}(d), a small peak at a positive shift of 140 \cm{} can be assigned to the $^B0^0_0$ decay. .
This excitation process is presumably due to collisions with hot reaction products before they are thermalized by the buffer gas. 
Lastly, a weak peak at $-1172$ \cm is attributed to the stretching mode $24^0_1$, while the strong peak at $-584$ \cm{} is assigned to CaF $A^2\Pi_{1/2} (v'=0) \rightarrow X^2\Sigma^+ (v''=1)$ decay. 

As seen in the relative heights of the $^A0_0^0$ and $^B0_0^0$ peaks in  Figs.~\ref{fig:Structure_2D}(d-e), both $\widetilde A \rightarrow \widetilde X$ and $\widetilde B \rightarrow \widetilde X$ transitions are promising for optical cycling as vibrational-state-preserving decays are strongly favored.
Because vibration-changing decays below the measurement sensitivity ($\sim 1\times10^{-3}$) or obscured by other peaks may not have been observed, the ratio of the intensities of these peaks is not strictly a VBR (see SI).
Therefore, to compare to theoretical calculations we plot (Figs. \ref{fig.data_plus_theory}(b-c)) the ratio of intensities of only the observed peaks.
In general, Figs.~\ref{fig.data_plus_theory}(b-c) show reasonable agreement between experiment and theory, other than the aforementioned underestimate of the decay to low-frequency bending modes.

\begin{figure}
    \centering
    \includegraphics[width = 0.95\textwidth]{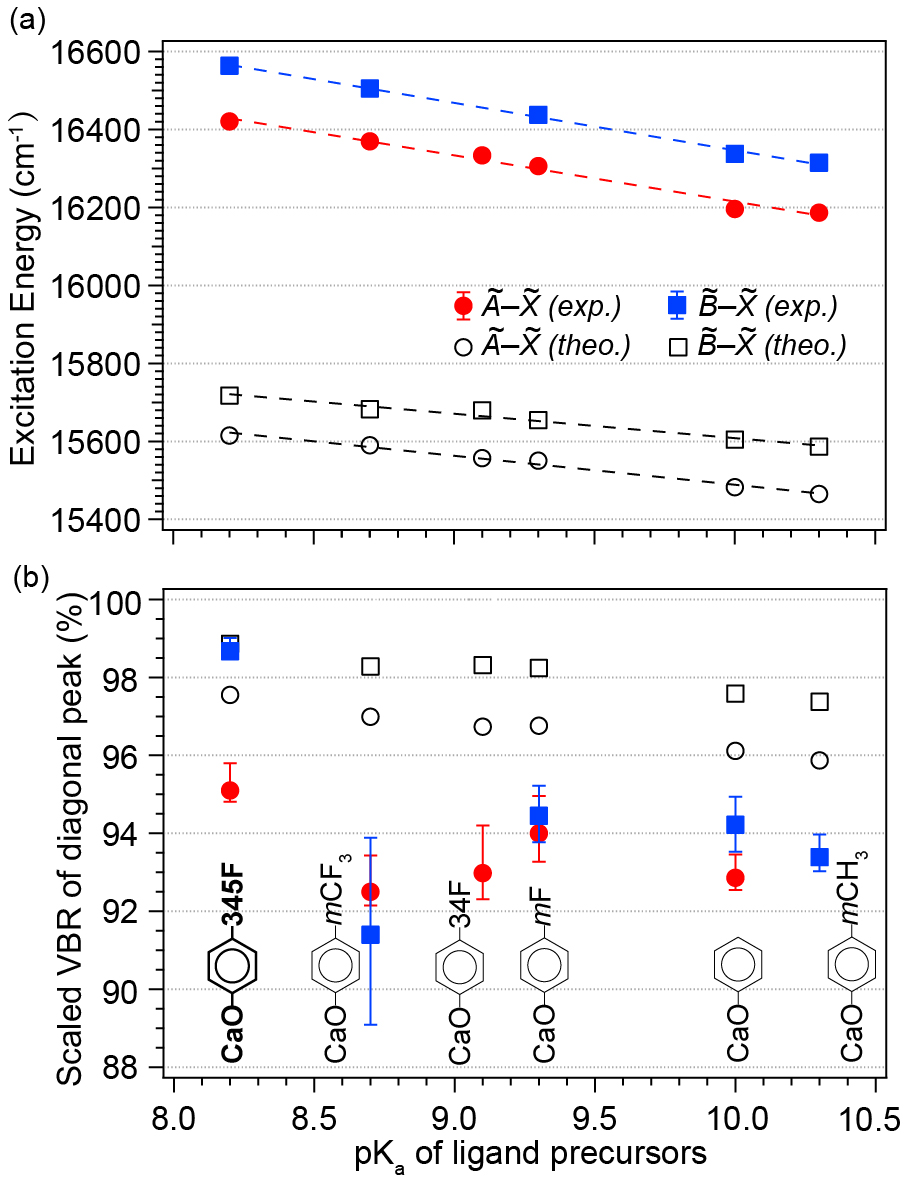}
    \caption{(a) $\widetilde A \leftarrow \widetilde X$ and $\widetilde B \leftarrow \widetilde X$ transition energies versus pK$_a$ for all molecular species studied here. 
    (b) Scaled VBR for diagonal decay as a function of pK$_a$. 
    } 
    \label{fig.pka}
\end{figure}

To explore the effect of the ligand on the OCC, Fig.~\ref{fig.pka} plots the measured transition energies and estimated $0_0^0$ VBRs as a function of the precursor pK$_a$ in solution. 
The pK$_a$ is a convenient parameter that indicates the strength of an acid R$-$OH, and therefore quantifies the electron-withdrawing capability of the R$-$O$^-$ ligand. 
A smaller pK$_a$ implies a more ionic Ca(I)$-$O bond in R$-$O$-$Ca.
As can be seen in Fig.~\ref{fig.pka}(a), the excitation energies follow a monotonic and apparently linear trend with pK$_a$. 
This behavior is qualitatively understood as an increase in HOMO-LUMO gap as the electron-withdrawing strength of the ligand increases \cite{mao2018HOMO}, yielding more localized molecular orbitals on the Ca atom (Fig. \ref{fig:Structure_2D}(b)) with a trend toward the atomic Ca$^+$ ion.

Figure \ref{fig.pka}(b) shows the estimated $0_0^0$ VBRs as a function of pK$_a$.
The $0^0_0$ VBR for each transition is estimated by normalizing the observed $0^0_0$ decay signal by the signal of all observed transitions plus an estimated contribution of the unobserved peaks (see SI). 
The error bars on each point represent the combination of the statistical standard error and the uncertainty from the unobserved peaks.
Systematic errors, discussed in the SI, are expected to be smaller than a few percent. 

Remarkably, across all six ligands and a considerable range of pK$_a$, the VBRs are relatively unchanged and always $\gtrsim 90$~\%, indicating that the OCC function can indeed be made orthogonal to the ligand molecule. 
The theoretical calculations in Fig.~\ref{fig.pka}(b) show an increase in VBR for stronger acids, as previously predicted~\cite{dickerson2021franck}.
This trend is consistent with the experimental data, and is understood as the localization of the electronic wavefunction on the Ca atom with a more ionic Ca(I)$-$O bond, leading to further isolation of the electronic and vibrational degrees of freedom \cite{dickerson2021franck}.
This suggests that while an OCC can be successfully attached to a wide range of molecules, performance may still be optimized by choosing ligands with strong electron withdrawing character.  

Together, these features illustrate that a quantum functional group, furnishing a means for gaining full quantum control of a wide range of molecules, should be possible. 
As an example, from the recorded structure and measured state lifetimes (see SI), we estimate a magneto-optical trap of CaO\phenylX{345F} is possible using the $\widetilde{B} - \widetilde{X}$ transition with 6-8 lasers (see SI)~\cite{augenbraun2020molecular}.
Interestingly, this is similar to the number of lasers required in molecules with roughly an order of magnitude fewer vibrational modes~\cite{Baum2021Establishing}.
Further, using the same laser system, single quantum states can be prepared via optical pumping and measured by state-resolved fluorescence with high fidelity. 

The ability to laser cool and prepare single quantum states of such `large' molecules opens the door to a host of new science.
The rich structures of these molecules allow robust encoding of quantum information \cite{albert2020robust} and provides new platforms for precision measurement and tests of fundamental physics \cite{augenbraun2020laser, hutzler2020polyatomic}
as well as quantum simulation and computing~\cite{blackmore2018ultracold, yu2019scalable}. 
Ultracold organic hydrocarbons can be produced via zero-energy photofragementation~\cite{lane2012ultracold} of those complexes and offer new opportunities for ultracold chemical reactions and collision studies~\cite{balakrishnan2016perspective}.  
Further, theoretical work suggests that the scheme demonstrated here can be continued to even larger molecules~\cite{dickerson2021optical} and extended to surfaces~\cite{guo2021surface}.

In summary, we have functionalized six precursor compounds (phenol, $m$-cresol, 3-fluorophenol,  3-(trifluoromethyl)phenol, 3,4,-bifluorophenol, and 3,4,5-trifluorophenol) with an optical-cycling center composed of a Ca(I)$-$O functional group.
The resulting molecules were studied at cryogenic temperature via 2D and dispersed fluorescence spectroscopy. 
The excitation energies of the molecules showed a linear correlation with the pK$_a$ of the organic precursors, providing a convenient way of discovering new molecules. 
Meanwhile, the vibrational branching ratios were largely unaffected by the choice of ligand and at a level sufficient for laser cooling and trapping as well as quantum state preparation and measurement.
This demonstration of the orthogonality of the OCC function to the ligand function lays the ground work for functionalizing molecules with quantum functional groups and establishes principles of chemical design that can be used to build molecules of increasing size, complexity, and function for quantum science and technology.

\emph{Acknowledgements --} The authors thank Timothy C. Steimle for sharing critical equipment and for useful discussions and also Nathaniel Vilas and Yicheng Bao for helpful discussions.
This work was supported in part by National Science
Foundation (Grants No. PHY-1255526, No. PHY-1415560, No. PHY-1912555, PHY-2110421, No. CHE-1900555, No. DGE-1650604, No. DGE-2034835, and No. OMA-2016245), Army Research Office (Grants No. W911NF-15-1-0121, No. W911NF-14-1-0378,
No. W911NF-13-1-0213,  W911NF-17-1-0071, and W911NF-19-1-0297) grants, and AFOSR (Grant No. FA9550- 20-1-0323).

\bibliographystyle{apsrev4-1_no_Arxiv}
\bibliography{allrefs}

\newpage 
~\newpage
\textbf{Supplementary Information}

\section{Spectroscopy system details}
The resulting CaO\phenylX{X} molecules were studied using two-dimensional (2D) spectroscopy via excitation and dispersed fluorescence~\cite{Reilly2006, Gascooke2011, Kokkin2014,augenbraun2021}, DLIF spectroscopy, and radiative decay. 
A tunable, pulsed optical parametric oscillator (OPO)
(with approximate parameters 5~\cm{} linewidth, 10~ns pulse duration, and 1~mJ pulse energy) illuminated the molecules at a delay of $\approx 1$~ms after the ablation pulse. 
The OPO wavelength was continuously tunable from 500~nm to 700~nm and the absolute wavelength was determined by a spectrum analyzer with a measurement accuracy of 0.5 nm. 
Molecular fluorescence was collected into a 0.67~m focal length Czerny-Turner style monochromator (numerical aperture $\approx \,$0.1) equipped with a 300 lines/mm grating (500~nm blaze).
The dispersed fluorescence was imaged onto a gated, intensified charge-coupled device camera (ICCD) cooled to $-30$~$^\circ$C. 
Given the system passband, a roughly $80$~nm wide spectral region of the DLIF could be recorded in a single image.

\section{Production of molecules}
The studied molecules were produced by the reaction of Ca atoms with ligand precursors in a cryogenic buffer-gas cell operated at $\approx$~9~K~\cite{hutzler2012buffer}.
Five precursor molecules -- phenol, $m$-cresol, 3-fluorophenol,  3-(trifluoromethyl)phenol and 3,4,5-trifluorophenol -- were purchased commercially and used without further purification. 
Gas-phase Ca atoms were introduced into the buffer gas by laser ablation of a metallic Ca target using the second-harmonic of a pulsed Nd:YAG laser (with approximate parameters of 20~mJ pulse energy, 100~$\mu$m spot size, 5~ns pulse length, and 10~Hz repetition rate).
The position of the ablation laser was continuously swept over the target with a moving mirror to avoid ablation-induced yield drifts.
A reservoir containing the ligand precursors was heated to a temperature between 300 K and 350 K to maintain a vapor pressure of 3-5 Torr (melting points of precursors given in Table~\ref{table:melting}). 
The resulting vapor was flowed into the cryogenic cell via a thermally isolated, heated fill line. 
The  calcium-bearing reaction products (CaO\phenylX{X}) were cooled via collisions with a He buffer gas of density $\approx 10^{15-16}$~cm$^{-3}$.

\section{Systematic Error}
In addition to the statistical uncertainty, there are several sources of systematic uncertainty (see Table~\ref{table:error}).
First, vibration changing decays with a strength below our experimental sensitivity (see Table~\ref{table:VBRs345f}-\ref{table:VBRscaoph&ch3}) can lead to a VBR error of up to 3\%.
Second, calibration error in the wavelength response of our instrument, including the imaging system, spectrometer, and ICCD camera, could lead to a VBR error of up to $\approx$ 1\%. 
Third, diagonal excitations from excited vibrational levels and subsequent decays can skew the measured VBRs if the excited states' VBRs differ from those of the ground state. 
Since the molecules are thermalized with the helium buffer gas, the population of excited vibrational levels is small 
and this error is estimated to be smaller than 0.5\%. 
Finally, the measured laser power fluctuations could lead to an error of up to 1\%. 
Further details are given in the SI.

\section{Laser Cooling Estimates}
Together, these features illustrate that a quantum functional group that furnishes a means for gaining full quantum control of a wide range of molecules should be possible. 
An estimate of the feasibility of such a scheme can be obtained by considering the number of lasers required to scatter on order 1000 photons from a molecule via the OCC; this number of scattered photons is typically enough to realize laser cooling and/or high-fidelity detection~\cite{mitra2020direct, augenbraun2020laser, Cheuk2018Lambda}. 
For example, for the $\widetilde{B} - \widetilde{X}$ transition in CaO\phenylX{345F}, three lasers to ``repump" the observed decays to $\nu_6, \nu_{24},$ and $\nu_{16}$ should be sufficient.
Additional lasers may be necessary to repump decays that change the rotational state of the molecule~\cite{augenbraun2020molecular}, and while the details depend on pending high-resolution spectroscopy, we estimate that in total roughly 6-8 lasers would be necessary to scatter $\sim$1000 photons. 
Interestingly, this is similar to the number of lasers required to achieve an equivalent photon budget in molecules with roughly an order of magnitude fewer vibrational modes~\cite{Baum2021Establishing, Zhang2021}.
Finally, given that we have measured the lifetime of these excited states to be roughly 20-30~ns (see SI), a strong scattering force can be expected, and techniques like laser cooling should be applicable.

\section{Electronic symmetry labels}
We label the electronic symmetries according to representations of the $C_{2v}$ point group, although for molecules with $C_s$ symmetry the appropriate symmetry species are $A'$ and $A''$.

\begin{table*}[h]
    \centering
    \caption{Observed and theoretical VBRs for all transitions studied in this work. Errors are standard error of fit and do not include systematic uncertainities. 
    }
    \renewcommand{\arraystretch}{1.2}
    \begin{tabular}{c c c c c}
    \hline
    \hline
    \multirow{2}{*}{Modes } &\multicolumn{4}{c}{CaOPh} \\
    \cline{2-5}
    & Exp. (A) & Theo. (A) & Exp. (B) & Theo. (B)  \\
    \cline{1-5}
    $0$ & $0.9339(29)$ & 0.9611 & $0.9446(68)$ & 0.9758\\
    $\nu_2$ & $0.0113(21)$ & $<10^{-4}$ & $0.0056(44)$ & $<10^{-4}$\\
    $\nu_4$ & $0.0468(15)$ & 0.0311 & $0.0419(29)$& 0.0185\\
    $\nu_9$ & $0.0058(11)$ & 0.0027 & $0.0033(29)$ & $0.0027$\\
    $\nu_{13}$ & $0.0023(8)$ & 0.0018 & $0.0037(27)$ & 0.0012 \\
    $\nu_{23}$ &  & 0.0017 & $0.0009(23)$ & 0.0008\\
    \hline
    \multirow{2}{*}{Modes } &\multicolumn{4}{c}{CaOPh-\textit{m}CH$_3$}  \\
    \cline{2-5}
    & Exp. (A) & Theo. (A) & Exp. (B) & Theo. (B) \\
    \cline{1-5}
    $0$ & & 0.9587 & $0.9387(32)$ & 0.9738 \\
    $\nu_6$  &  & 0.0175 & $0.0232(17)$ & 0.0110\\
    $\nu_7$ &  & 0.0151 & $0.0294(16)$ & 0.0090 \\
    $\nu_{12}$ &  & 0.0024 & $0.0046(16)$ & 0.0024 \\
    $\nu_{14}$ &  & 0.006 & $0.0041(16)$ & 0.0011 \\
    \hline
    \multirow{2}{*}{Modes } &\multicolumn{4}{c}{CaOPh-\textit{m}F}  \\
    \cline{2-5}
    & Exp. (A) & Theo. (A) & Exp. (B) & Theo. (B) \\
    \cline{1-5}
    $0$ & $0.9464(71)$ & 0.9676 & $0.9501(53)$ & 0.9824\\
    $\nu_1$ & $0.0037(15)$ & 0.0009 & $0.0065(37)$ & $<10^{-4}$\\
    $\nu_5$ & $0.0314(16)$ & 0.0222 & $0.0282(23)$ & 0.0122\\
    $\nu_6$ & $0.0107(53)$ & 0.0031 & $0.0090(23)$ & 0.0013\\
    $\nu_{10}$ & $0.0048(13)$ & 0.0018 & $0.0062(23)$ & 0.0018\\
    $\nu_{14}$ & $0.0030(12)$ & 0.0010 & & 0.0007\\
    \hline
    \multirow{2}{*}{Modes } &\multicolumn{4}{c}{CaOPh-\textit{m}CF$_3$}  \\
    \cline{2-5}
    & Exp. (A) & Theo. (A) & Exp. (B) & Theo. (B) \\
    \cline{1-5}
    $0$ & $0.9339(28)$ & 0.9699 & $0.9308(184)$ & 0.9828\\
    $\nu_2$ & $0.0137(13)$ & 0.0028 & $0.0144(74)$ & 0.0003\\
    $\nu_5$ & $0.0018(11)$ & 0.0006 & & 0.0004 \\
    $\nu_7$ & $0.0320(11)$ & 0.0154 & $0.0343(81)$ & 0.0076\\
    $\nu_9$ & $0.0073(10)$ & 0.0031 & $0.0091(55)$ & 0.0002\\
    $\nu_{10}$ & $0.0050(10)$ & 0.0018 & & 0.0008\\
    $\nu_{15}$ & $0.0039(10)$ & 0.0013 & $0.0084(52)$ & 0.0014\\
    $\nu_{23}$ & $0.0025(10)$ & 0.0009 & $0.0030(49)$ & 0.0008\\
    \hline
    \multirow{2}{*}{Modes } &\multicolumn{4}{c}{CaOPh-34F}  \\
    \cline{2-5}
    & Exp. (A) & Theo. (A) & Exp. (B) & Theo. (B) \\
    \cline{1-5}
    $0$ & $0.9409(51)$ & 0.9674 && 0.9831\\
    $\nu_2$ & $0.0050(35)$ & 0.0012 && $<10^{-4}$\\
    $\nu_5$ & $0.0341(32)$ & 0.0136 && 0.0073\\
    $\nu_6$ & $0.0199(32)$ & 0.0112 && 0.0050\\
    \hline
    \multirow{2}{*}{Modes } &\multicolumn{4}{c}{CaOPh-345F}  \\
    \cline{2-5}
    & Exp. (A) & Theo. (A) & Exp. (B) & Theo. (B) \\
    \cline{1-5}
    $0$ & $0.9576(23)$ & 0.9755 & $0.9899(12)$ & 0.9886\\
    $\nu_1$ & $0.0046(12)$ & $<10^{-4}$ && $<10^{-4}$\\
    $\nu_6$ & $0.0270(10)$ & 0.0159 & $0.0064(7)$ & 0.0072\\
    $\nu_8$ & $0.0058(10)$ & 0.0031 && 0.0007\\
    $\nu_{16}$ & $0.0031(9)$ & 0.0018 & $0.0014(7)$ & 0.0010\\
    $\nu_{24}$ & $0.0020(9)$ & 0.0011 & $0.0023(6)$ & 0.0003\\
    \hline
    \hline
    \end{tabular}
    \label{table:VBRs&FCFs}
\end{table*}

\section{Measured VBRs}

We summarize the measured VBRs of each molecular species in Table \ref{table:VBRs&FCFs}. The 2D and DLIF spectra of all other species are illustrated in Figure \ref{fig.caoph-all-dlif} and discussed in supplementary material. The vibrational frequencies of all resolved fundamental modes are summarized in Table \ref{table:vib.freq.} and the corresponding vibrational displacements are given in Fig. \ref{fig:vib.modes}. The theoretical frequencies of all modes in \X~state are listed in Table \ref{table:freqall} and all theoretical VBRs ($>10^{-4}$) are summarized in Tables \ref{table:VBRs345f} - \ref{table:VBRscaoph&ch3}.

\subsection{Theoretical calculations of vibrational frequencies and FCFs of all molecules}
All molecular optimized geometries and frequency calculations were obtained on a superfine grid in Gaussian16 at the PBE0-D3/def2-tzvppd level of theory using density functional theory/time-dependent density functional theory methods \cite{perdew1996rationale,grimme2010consistent,rappoport2010property,frisch2016gaussian}.  Franck-Condon factors were calculated within the parallel harmonic approximation with ezFCF \cite{gozem2021ezspectra}.  Due to the isolated density of the optical cycling centers and rigidity of the molecule, the parallel harmonic approximation is expected to be reliable.   However, anharmonicity effects and potential errors were discussed further in a previous paper \cite{dickerson2021franck}.  Additionally, Franck-Condon factors calculated with DFT missed vibronic coupling effects, due to the single reference nature of the DFT formalism.  Higher-level theory is needed to express vibronic coupling effects more accurately, but we find DFT is a suitable qualitative approximation to predict FCFs for these molecules.  The theoretical VBRs in Table \ref{table:VBRs345f}-\ref{table:VBRscaoph&ch3} are converted by the calculated FCFs with the formula \cite{augenbraun2020molecular}:
\begin{align}
    b_{iv',fv''} & = \frac{I_{iv',fv''}}{\sum_{fv''}I_{iv',fv''}} = \frac{A_{iv',fv''}}{\sum_{fv''}A_{iv',fv''}}\nonumber\\
    & = \frac{|\mu_{iv',fv''}|^2~\times~(\nu_{iv',fv''})^3}{\sum_{fv''} |\mu_{iv',fv''}|^2~\times~(\nu_{iv',fv''})^3}\nonumber\\
    &\approx \frac{\text{FCF}_{iv',fv''}~\times~ \nu^3_{iv',fv''} }{\sum_{fv''} \text{FCF}_{iv',fv''} ~\times~\nu^3_{iv',fv''}}
\end{align}

\noindent where $i$ and $f$ indicate the initial and final states, respectively. $b_{iv',fv''}$ is the branching ratio, $I_{iv',fv''}$ is the intensity, $A_{iv',fv''}$ is the Einstein coefficient for spontaneous emission, $\mu_{iv',fv''}$ is the transition dipole moment and $\nu_{iv',fv''}$ is the transition frequency.  
\subsection{2D and DLIF of other molecules}
The 2D and DLIF spectra of other five molecules, including CaOPh, CaOPh-\textit{m}CH$_3$, CaOPh-\textit{m}F, CaOPh-\textit{m}CF$_3$ and CaOPh-34F are presented in Figure \ref{fig.caoph-all-dlif}. The 2D spectra were recorded by monitoring the emission wavelength when scanning the excitation wavelength at a step size of 0.2 nm at the region of $600-620$ nm. Besides the transition bands from the molecular species of interests, indicated by the green dotted lines, those from unwanted species of Ca, CaOH and CaF, indicated by the orange dash lines, are also observed. The Ca atom is directly produced by laser ablation while CaOH and CaF are formed by the reactions with water or precursor ligands. By putting the laser wavelengths at the excitation bands of CaO\phenylX (green dotted lines), the respective DLIF spectra are obtained for both $\widetilde A \rightarrow \widetilde X$ and $\widetilde B \rightarrow \widetilde X$  transitions. All spectra are fitted with Pearson functions (red traces) to extract VBRs. In comparison to the theoretical vibrational frequencies (Table \ref{table:freqall}) and VBRs (Tables \ref{table:VBRs345f}-\ref{table:VBRscaoph&ch3}), the resolved vibrational peaks can be assigned to different vibrational modes, as labeled in the spectra. 

Figures \ref{fig.caoph-all-dlif}(a) shows the 2D spectrum of CaOPh. Besides the two bands from CaOPh indicated by the green dotted lines, several bands are observed for Ca and CaOH. The strong CaOH band at around 617 nm is overlapped with the CaOPh $\widetilde A - \widetilde X$ band, which results in many CaOH peaks, as denoted by *, in the dispersed spectrum of CaOPh in Fig. \ref{fig.caoph-all-dlif}(b). The origin peak, labeled as $^A0^0_0$, represents the fluorescence decay from excited \A~$(v'=0)$ state to the ground \X~ $(v''=0)$ state. The shifts and relative intensities of peaks at $-312$~\cm and $-627$~\cm agree well with simulations and they can be assigned to Ca-O stretching modes $\nu_4$ (314 \cm) and $\nu_9$ (631 \cm), respectively. The weak peak at $-878$ \cm is assigned to $\nu_{13}$ (903 \cm), while a shelf peak at $-47$ \cm is from the bending mode $\nu_1$ (61 \cm). The peak at positive shift of 117 \cm is ascribed to the $\widetilde B (v'=0) \rightarrow \widetilde X (v''=0)$ decay. The presence of the excited $\widetilde B$ state when exciting to the $\widetilde A$ state has been observed for all molecules, which is likely due to the buffer-gas collision-induced excitation. Without the contamination of CaOH, the $\widetilde B \rightarrow \widetilde X$ spectrum of CaOPh in Fig. \ref{fig.caoph-all-dlif}(c) is straightforward to assign. The same vibrational modes are resolved and the respective peaks are labeled as $^B1^0_1$, $^B4^0_1$, $^B9^0_1$ and $^B13^0_1$. Besides, the shift of the peaks labeled as $^A0^0_0$ at $-138$ \cm, not matching frequencies of any vibrational modes, is due to the collision-induced relaxation of $\widetilde B (v'=0) \rightarrow \widetilde A (v'=0) $, followed by the fluorescence decay to $\widetilde X (v''=0)$.

The DLIF spectra of CaOPh-\textit{m}CH$_3$ are displayed in Figs. \ref{fig.caoph-all-dlif}(e) and (f). The excitation wavelength of $\widetilde A \leftarrow \widetilde X$ at 617.80 nm can also excite the CaOH transition $\widetilde A^2\Pi_{1/2} (02^20) \leftarrow \widetilde X^2\Sigma^+$ (100), which results in a broad peak, labeled as *, at around $-100$~\cm and contributes to the diagonal peak. The difficulty to subtract the CaOH contribution makes it impossible to estimate the VBR for $\widetilde A \rightarrow \widetilde X$ of CaOPh-\textit{m}CH$_3$. The two peaks at $-285$~\cm~and~$-327$~\cm\ are assigned to the stretching modes $\nu_6$ and $\nu_7$, respectively, in comparison to the theoretical frequencies (Table \ref{table:freqall}) and relative VBRs (Table \ref{table:VBRscaoph&ch3}). The $\widetilde B \rightarrow \widetilde X$ is more complicated due to the relaxation to $\widetilde A$ and the overlapped Ca transition lines. Four vibrational peaks have been observed and labeled as $^B6^0_1$, $^B7^0_1$, $^B12^0_1$ and $^B14^0_1$, which are predicted to be the top four off-diagonal modes. The relaxation to $\widetilde A$ are confirmed with the observed peaks of $^A0^0_0$, $^A6^0_1$ and $^A7^0_1$. The plus symbol $^+$ labels the peaks due to decays from Ca $^3S_1$, generated by laser ablation, to Ca $^3P_J (J=0,1,2)$. The VBR calculations have subtracted the contribution from Ca emissions.

Figure \ref{fig.caoph-all-dlif}(g) illustrates the 2D spectrum of CaOPh-\textit{m}F. Besides the strong bands from CaF and CaOPh-\textit{m}F, there are two weak bands from CaOPh, which is the by-product of the reaction between Ca and 3-fluorophenol. The dispersed spectrum obtained by pumping the $\widetilde A \leftarrow \widetilde X$ transition resolves five off-diagonal vibrational modes, including one bending mode $\nu_1$ and four stretching modes $\nu_5$, $\nu_6$, $\nu_{10}$ and $\nu_{14}$. The higher intensity of peak $^A6^0_1$ than peak $^A5^0_1$, which is contrary to the theoretical VBRs, is due to the contribution of CaOH $\widetilde A^2\Pi_{1/2} (010) \rightarrow \widetilde X^2\Sigma^+$ (010), as labeled by symbol *. CaOH $\widetilde A^2\Pi_{1/2} (010) \rightarrow \widetilde X^2\Sigma^+$ (000)  also has a weak contribution to the diagonal peak. In Fig. \ref{fig.caoph-all-dlif}(i), four same vibrational peaks, labeled as $^B1^0_1$, $^B5^0_1$, $^B6^0_1$ and $^B10^0_1$, have been observed for the $\widetilde B \rightarrow \widetilde X$ transition and agree perfectly with the predictions (blue lines). The additional relaxation peak $^A0^0_0$ is also observed without surprise. 

The spectra of CaOPh-\textit{m}CF$_3$ are more complicated due to the non-planar structure introduced by the substituent of CF$_3$ group. By comparing to the theoretical frequencies (Table \ref{table:freqall}) and VBRs (Table \ref{table:VBRscaoph&ch3}), seven vibrational peaks have been assigned and labeled in the dispersed spectrum of $\widetilde A \rightarrow \widetilde X$ in Fig.\ref{fig.caoph-all-dlif}(k). For the $\widetilde B \rightarrow \widetilde X$ decay in in Fig. \ref{fig.caoph-all-dlif}(l), five vibrational peaks are assigned as $^B2^0_1$, $^B7^0_1$, $^B9^0_1$, $^B15^0_1$ and $^B23^0_1$, while five additional peaks are observed for the $\widetilde A \rightarrow \widetilde X$ transition due to relaxation from $\widetilde B$. The CaF peak from the transition of $A\,^2\Pi_{3/2} (v'=0)\rightarrow X\,^2\Sigma^+ (v''=1)$ is also observed in coincidence. The CaF intensity in the diagonal peak is subtracted when calculating the VBR.  

The reaction of 3,4,5-trifluorophenol with Ca has produced CaOPh-34F as a secondary product, as confirmed by the band of $\widetilde A \rightarrow \widetilde X$ transition at 612.04 nm in Fig. \ref{fig.caoph-all-dlif}(m). The corresponding spectrum has only resolved three peaks. The diagonal peak labeled as $^A0^0_0$ represents the fluorescence decay from excited \A~$(v'=0)$ state to the ground \X~ $(v''=0)$ state. Two vibrational peaks with relative shifts of $-275$~\cm~and~$-291$~\cm\ agree well with 
the theoretical frequencies of Ca-O stretching modes $\nu_5$ (278 \cm) and $\nu_6$ (293 \cm), as listed in Table \ref{table:freqall}, respectively. The relative intensites of these two peaks are also in good agreement with theoretical VBRs (blue lines, Table \ref{table:VBRs345f}). 

The frequencies of all resolved vibrational modes are summarized in Table \ref{table:vib.freq.} and the corresponding vibrational displacements are given in Fig. \ref{fig:vib.modes}.

\subsection{Curve fitting and error analysis}
We start by correcting the DLIF spectra obtained for a fixed excitation wavelength to the wavelength response of the system composed of the spectrometer, the iCCD camera and associated optics. The wavelength response was obtained for an almost identical system by using a tungsten lamp as a source of blackbody radiation and a pyrometer to measure the temperature. The collected camera images were first averaged, then integrated along the slit axis and finally corrected for the wavelength response. Subsequently, the background images with excitation light scatter waere similarly processed and subtracted from the data to obtain the DLIF spectra shown in Fig. \ref{fig.caoph-all-dlif}. 

The shape of individual peaks in the DLIF spectra are representative of the instrument response function. The lineshape of peaks are slightly skewed due to imperfect alignment of the optical axis. We confirm that the lineshape is identical for molecular decay lines, atomic decay lines and lines due to scattered excitation light, indicating that the lineshape is uniquely defined by the instrument response function. In order to extract the area under each peak and also be able to resolve peaks separated by less than the spectrometer resolution, we need to fit each peak. Since the lineshape of the peaks are arbitrary and do not fit well to Gaussian or Lorentzian distributions, we choose to fit the peaks to a Pearson distribution. A Pearson distribution is a useful tool to fit arbitrary lineshapes leveraging the first four non-zero moments - mean, variance, skewness and kurtosis. In addition, we notice that the peak centers and tails fit to two different distributions. Since the only purpose of the fit is to accurately extract the area under each peak, we fit each peak to a sum to two Pearson distributions - one for the centers and one for the tails. We find that the experimental data fits optimally to two, Type-$IV$ Pearson distributions defined as 

\begin{equation}
    f(x) \propto \left[1+\frac{x^2}{\alpha^2}\right]^{-m}\times\text{exp}\left[-\nu \, \text{tan}^{-1}\left(\frac{x}{\alpha}\right)\right],
\end{equation}
where the parameters $m$, $\alpha$ and $\nu$ reflect the moments. We estimate start parameters by isolating and fitting the main peak near 0~\cm. Next, we fit the entire spectrum simultaneously to the sum of $N$ sets of 
Pearson distributions, each having a variable position and amplitude but identical width. This width is defined by 8 parameters that define the moments of the center and tail distributions. In all, the total spectra containing $N$ peaks, is fully fit to $2N+8$ free parameters. The result of the fit is superimposed with the DLIF spectra in Fig. \ref{fig.caoph-all-dlif}.

The intensity of the fit peaks is simply extracted as the fit amplitudes. This intensity is also proportional to the area enclosed under the peaks since all peaks, by definition, have the same width. Hence the intensity ratio, or amplitude ratio, is directly related to the VBR. We obtain the statistical fit uncertainty in the estimation of the VBR from the covariance matrix of the fit parameters. This statistical error is depicted in Fig. \ref{fig.data_plus_theory}. We note that using a different model like a Gaussian distribution for all peak fitting leads to VBRs that differ by at most 1\%. We count that as a systematic error from the fit model. In addition, we take into account other sources of systematic error in your measurements. These sources are summarized in Table \ref{table:error}. 

Since the true VBR of the decay relies on the total contributions of all possible decay pathways, the primary source of systematic error is the unknown intensity of any unobserved decay. We estimate this contribution of unobserved peaks as follows. We first calculate a scaling factor $C$ by averaging the ratio of theoretical VBR ($T_i$) to experimental VBR ($S_i$) for all observed vibrational peaks:
\begin{equation}
\begin{aligned}
    C = \frac{1}{p}\sum_{\substack{i=0}}^{p} \frac{T_{i}}{S_{i}}.
\end{aligned}
\end{equation}
The scaled diagonal VBR ($S'_0$) is calculated as
\begin{equation}
\begin{aligned}
  S'_{0}=\frac{S_{0}}{\sum_{\substack{i=0}}^{p}S_{i} + \sum_{\substack{i=p+1}}^{N}\frac{T_{i}}{C}},
\end{aligned}
\end{equation}
where the observed diagonal VBR ($S_0$) is divided by the sum of all observed VBRs and the scaled unobserved VBRs. When excluding the diagonal peak in equation (3), we can obtain the lowest scaled VBR $S''_0$. While the scaled VBR $S'_0$ is used as the practical experimental point in Fig. \ref{fig.pka}b, the measured VBR $S_0$ and the new scaled VBR $S''_0$ are used as the upper and lower bounds, respectively, of the uncertainties due to the unobserved peaks. As shown in Table \ref{table:unobs-error}, the uncertainties are in the range of $0.34\% - 3.08\%$.

The next considerable source of systematic error is expected to be the uncertainty associated with the imperfect calibration of the wavelength response of the instrument. All DLIF spectra have been calibrated using a sensitivity curve at the whole wavelength, which is measured with a tungsten lamp as a blackbody source. An inaccurate measurement of filament temperature or contamination of the filament would cause the erroneous of the sensitivity response. Assuming a 10\% uncertainty in the sensitivity at the wavelength range of our spectra (typically 20 nm covering the diagonal peak and the stretching-mode peak), We can estimate that the final VBR is off by around 0.5\%. Considering the use of slightly different mirror types and AR coated lenses in our experiment from the measurement of the sensitivity curve, the systematic error would be added up to $\approx$ 1\%. 

Next, we consider the effect of ``diagonal" excitation of vibrationally excited modes in the $\widetilde{X}$ manifold. Since our excitation light is 0.2~nm wide, it can simultaneously excite $\widetilde{X} (v''=0) \rightarrow \widetilde{A} (v'=0)$, $\widetilde{X} (v''=1) \rightarrow \widetilde{A} (v'=1)$, $\widetilde{X} (v''=2) \rightarrow \widetilde{A} (v'=2)$, etc. Table \ref{table:diagonalVBR} shows the theoretical VBRs from the different excited vibrational levels of $\widetilde{A}$ state of CaOPh. The VBRs from the $\widetilde{A} (v'=1)$ of the bending mode $\nu_1$ are almost identical to those from $\widetilde{A} (v'=0)$, yielding almost no change of our VBRs in the diagonal excitation. However, VBRs from the $\widetilde{A} (v_4'=1)$ of the stretching mode $\nu_4$ differ significantly from those associated with $\widetilde{A} (v'=0)$ . The diagonal decay is predicted to be 0.8963, smaller than the value of 0.9611 associated with the diagonal VBR from $\widetilde{A} (v'=0)$ decay. Meanwhile, the strongest off-diagonal decay is 0.0598 from $\widetilde{A} (v_4'=1)$, larger than the value of 0.0311 from $\widetilde{A} (v'=0)$ decay. Because we didn't observe the blue-detuned peak $\widetilde{A} (v_4'=1) \rightarrow \widetilde{X} (v''=0)$ with a theoretical VBR of 0.0354 above the noise level of $10^{-3}$, the thermal population of the first vibrational state of the stretching mode $\nu_4$ is estimated to be at most about 3\%. This would yield an 0.2\% difference of the diagonal VBR. Considering the VBRs differences from other vibrational modes and other molecules, we estimate that these unwanted diagonal excitations can contribute at most 0.5\% to the intensity of the main peak.    

The final source of systematic uncertainty that we consider is the effect of scattered excitation light impinging on the detector. Although the start of the camera gate pulse is delayed from the OPO excitation pulse (10~ns wide) by $\approx$30~ns, we still observe some scattered light from the excitation pulse in the camera images. This scattering can be eliminated by delaying the gate pulse farther but since the excited state lifetime is $\approx$25~ns, we collect fewer decay photons which reduces our signal. Instead, we subtract the excitation light by separately measuring only the scattered light after taking every DLIF spectra over 5,000 repetitions. Typically, the OPO pulse intensity is about 1\% up to 50\% of the molecular fluorescence signal and can vary by up to $\approx 20\%$ within that duration. This can lead to up to a 0.5\% error. The total of all these sources, added in quadrature, leads to a nominal estimate of systematic uncertainty of $ 1.62\% - 3.48\%$.      

\renewcommand*{\thetable}{S\arabic{table}}
\setcounter{table}{0} 
\begin{table*}
\centering
\begin{tabular}{c|c c c c c c c}
\hline
\hline
Molecules & PhOH-mCH$_3$ & Phenol  & PhOH-3F & PhOH-34F & PhOH-mCF$_3$  & PhOH-35F & PhOH-345F \\ \hline
Melting Point ($^o$C) & 12.2(3) & 40.89(1)  & 14(1) & $34-38$ & $-0.9$  & $54-58$ & 57 \\ 
pKa\cite{hanoian2010hydrogen} & 10.3 & 10.0  & 9.3 & 9.1 & 8.7  & 8.4 & 8.2 \\
Hammett parameter\cite{dickerson2021franck, hammett1937effect} & $-0.07$ & 0  & 0.34 & 0.40 & 0.43  & 0.67 & 0.74  \\
\hline
\hline
\end{tabular}
\caption{Melting points, pKa and Hammett parameters for the different molecular species described in this work. The pKa and Hammett parameter can be linked by the derived Hammett equation \cite{hammett1937effect}: pKa (X) = pKa (H) - $\sigma \rho$,  where pKa (X) = $-$ log K (X), pKa (H) = $-$ log K (H), K (X) and K (H) are the equilibrium constants for a substituted species and unsubstituted phenol, respectively. $\sigma$ is the Hammett parameter and $\rho$ is the reaction constant.}
\label{table:melting}
\end{table*}

\begin{table*}
    \centering
    \setlength{\tabcolsep}{5pt}
    \renewcommand{\arraystretch}{1}
\begin{tabular}{c c c c c c }
\multicolumn{6}{c}{X-A Transition}  \\
\hline
\hline
Molecules & Measured VBR & Scaled VBR   & Scaled VBR exluding  & Upper error bar & Lower error bar \\
& $S_0$& $S'_0$ & main peak $S''_0$ & ($S_0 - S'_0$)& ($S'_0 - S''_0)$\\
\hline
CaOPh-mCH3 &        &        &        &        &        \\
CaOPh      & 0.9339 & 0.9286 & 0.9274 & 0.0053 & 0.0012 \\
CaOPh-mF   & 0.9464 & 0.9400 & 0.9383 & 0.0064 & 0.0017 \\
CaOPh-34F  & 0.9409 & 0.9298 & 0.9254 & 0.0111 & 0.0044 \\
CaOPh-mCF3 & 0.9339 & 0.9250 & 0.9230 & 0.0089 & 0.0021 \\
CaOPh-345F & 0.9576 & 0.9510 & 0.9493 & 0.0066 & 0.0018 \\
\hline
\hline
\\
\multicolumn{6}{c}{X-B Transition}  \\
\hline
\hline
Molecules & Measured VBR & Scaled VBR   & Scaled VBR exluding & Upper error bar & Lower error bar \\
& $S_0$& $S'_0$ & main peak $S''_0$ & ($S_0 - S'_0$)& ($S'_0 - S''_0)$\\
\hline
CaOPh-mCH3 & 0.9387 & 0.9339 & 0.9323 & 0.0048 & 0.0016 \\
CaOPh      & 0.9446 & 0.9422 & 0.9412 & 0.0024 & 0.0010 \\
CaOPh-mF   & 0.9501 & 0.9445 & 0.9403 & 0.0056 & 0.0042 \\
CaOPh-34F  &        &        &        &  &  \\
CaOPh-mCF3 & 0.9308 & 0.9140 & 0.9000 & 0.0168 & 0.0140 \\
CaOPh-345F & 0.9899 & 0.9867 & 0.9866 & 0.0032 & 0.0002 \\
\hline
\hline
\end{tabular}
\caption{The comparisons of measured diagonal VBR and scaled diagonal VBR. The differences indicate the systematic uncertainties of unobserved vibrational peaks. 
}
\label{table:unobs-error}
\end{table*}

\renewcommand*{\thetable}{S\arabic{table}}
\begin{table*}
\centering
\begin{tabular}{c|c}
\hline
\hline
Error source & Percentage \\ \hline

Unobserved peaks & $0.34\%-3.08\%$  \\
Instrument wavelength response & $\approx1\%$\\
Diagonal excitations from excited vibrational levels & $<0.5\%$ \\
OPO power fluctuation & $<0.5\%$\\
Fitting model & $<1\%$\\
\hline
Totol error & $1.62\%-3.48\%$\\
\hline
\hline
\end{tabular}
\caption{Summary of systematic error sources in the DLIF measurement.}
\label{table:error}
\end{table*}

\begin{table*}
    \centering
    \setlength{\tabcolsep}{7pt}
    \renewcommand{\arraystretch}{1.2}
    \begin{tabular}{c c c c c c c}
    \hline
    \hline
    \multirow{2}{*}{Vib. modes } &\multicolumn{2}{c}{CaOPh} && \multirow{2}{*}{Vib. modes } &\multicolumn{2}{c}{CaOPh-mCH$_3$}\\
    \cline{2-3}\cline{6-7} & Exp.& Theo.& & & Exp.& Theo.    \\
    \cline{1-3}\cline{5-7}
    $\nu_2$ & 44(6) & 61 & &$\nu_6$ & 285(2) & 291\\
    $\nu_4$ & 312(2) & 314 & &$\nu_7$ & 327(2) & 329\\
    $\nu_9$ & 627(6) & 631 & &$\nu_{12}$ & 636(4) & 640\\
    $\nu_{13}$ & 878(6) & 903 & &$\nu_{14}$ & 775(5) & 790\\
    $\nu_{23}$ & 1304(4) & 1347 & & & & \\
    \hline
  \multirow{2}{*}{Vib. modes } &\multicolumn{2}{c}{CaOPh-mF} && \multirow{2}{*}{Vib. modes } &\multicolumn{2}{c}{CaOPh-mCF$_3$}\\
    \cline{2-3}\cline{6-7} & Exp.& Theo.&&& Exp. & Theo.\\
    \cline{1-3}\cline{5-7}
    $\nu_1$  & 47(15) & 55 & &$\nu_2$ & 51(10) & 46\\
    $\nu_5$ & 290(2) & 297 & &$\nu_5$ & 182(12) & 183\\
    $\nu_6$ & 355(4) & 359 & &$\nu_7$ & 277(3) & 280\\
    $\nu_{10}$ & 625(4) & 630 & &$\nu_9$ & 345(2) & 350\\
    $\nu_{14}$ & 788(4) & 802 & &$\nu_{10}$ & 379(2) & 381\\
    & & & &$\nu_{15}$ & 626(9) & 634\\
    & & & &$\nu_{23}$ & 942(4) & 962\\
    \hline
      \multirow{2}{*}{Vib. modes } &\multicolumn{2}{c}{CaOPh-34F} && \multirow{2}{*}{Vib. modes } &\multicolumn{2}{c}{CaOPh-345F}\\
    \cline{2-3}\cline{6-7} & Exp.& Theo.&&& Exp.& Theo.    \\
    \cline{1-3}\cline{5-7}
    $\nu_2$ & 58(7) & 54 & &$\nu_1$ & 51(3) & 51\\
    $\nu_5$ & 275(3) & 278 & &$\nu_6$ & 266(3) & 272\\
    $\nu_6$ & 292(3) & 293 & &$\nu_8$ & 309(2) & 315\\
    &   & & &$\nu_{16}$ & 663(6) & 686\\
    &   & & &$\nu_{24}$ & 1173(6) & 1210\\
    \hline
    \hline
    \end{tabular}
    \caption{Comparison of the observed and calculated frequencies for resolved fundamental vibrational modes of all species studied in this work. Values are given in units of \cm. }
    \label{table:vib.freq.}
\end{table*}

\begin{table*}
    \centering
    \setlength{\tabcolsep}{5pt}
    \renewcommand{\arraystretch}{1.5}
    \begin{tabular}{c c c c c c c c c c c c c c c c c}
    \hline
    \hline
    \multicolumn{2}{c}{CaOPh-\textit{m}CH$_3$} & & \multicolumn{2}{c}{CaOPh} & & \multicolumn{2}{c}{CaOPh-\textit{m}F}&&\multicolumn{2}{c}{CaOPh-34F} & & \multicolumn{2}{c}{CaOPh-\textit{m}CF$_3$} & & \multicolumn{2}{c}{CaOPh-345F}\\
    \cline{1-2}\cline{4-5}\cline{7-8}\cline{10-11}\cline{13-14}\cline{16-17}
    Modes & Freqs.&  & Modes & Freqs. &  & Modes & Freqs. &&Modes & Freqs. &  & Modes & Freqs. &  & Modes & Freqs.\\
    \cline{1-2}\cline{4-5}\cline{7-8}\cline{10-11}\cline{13-14}\cline{16-17}
    $\nu_1$    & 37  & & $\nu_1$    & 61  & & $\nu_1$    & 55  & & $\nu_1$    & 51  & & $\nu_1$    & 11  & & $\nu_1$  & 50  \\
    $\nu_2$    & 55  & & $\nu_2$    & 61  & & $\nu_2$    & 59  & & $\nu_2$    & 54  & & $\nu_2$    & 46  & & $\nu_2$  & 51  \\
    $\nu_3$    & 59  & & $\nu_3$    & 247 & & $\nu_3$    & 238 & & $\nu_3$    & 166 & & $\nu_3$    & 58  & & $\nu_3$  & 147 \\
    $\nu_4$    & 208 & & $\nu_4$    & 314 & & $\nu_4$    & 249 & & $\nu_4$    & 241 & & $\nu_4$    & 128 & & $\nu_4$  & 222 \\
    $\nu_5$    & 247 & & $\nu_5$    & 426 & & $\nu_5$    & 297 & & $\nu_5$    & 278 & & $\nu_5$    & 183 & & $\nu_5$  & 260 \\
    $\nu_6$    & 291 & & $\nu_6$    & 448 & & $\nu_6$    & 359 & & $\nu_6$    & 293 & & $\nu_6$    & 246 & & $\nu_6$  & 272 \\
    $\nu_7$    & 329 & & $\nu_7$    & 529 & & $\nu_7$    & 471 & & $\nu_7$    & 366 & & $\nu_7$    & 280 & & $\nu_7$  & 277 \\
    $\nu_8$    & 457 & & $\nu_8$    & 629 & & $\nu_8$    & 502 & & $\nu_8$    & 378 & & $\nu_8$    & 331 & & $\nu_8$  & 315 \\
    $\nu_9$    & 480 & & $\nu_9$    & 631 & & $\nu_9$    & 523 & & $\nu_9$    & 471 & & $\nu_9$    & 350 & & $\nu_9$  & 363 \\
    $\nu_{10}$ & 527 & & $\nu_{10}$ & 714 & & $\nu_{10}$ & 630 & & $\nu_{10}$ & 486 & & $\nu_{10}$ & 381 & & $\nu_{10}$ & 364 \\
    $\nu_{11}$ & 594 & & $\nu_{11}$ & 782 & & $\nu_{11}$ & 642 & & $\nu_{11}$ & 590 & & $\nu_{11}$ & 470 & & $\nu_{11}$ & 604 \\
    $\nu_{12}$ & 640 & & $\nu_{12}$ & 837 & & $\nu_{12}$ & 706 & & $\nu_{12}$ & 604 & & $\nu_{12}$ & 477 & & $\nu_{12}$ & 518 \\
    $\nu_{13}$ & 715 & & $\nu_{13}$ & 903 & & $\nu_{13}$ & 781 & & $\nu_{13}$ & 646 & & $\nu_{13}$ & 536 & & $\nu_{13}$ & 589 \\
    $\nu_{14}$ & 790 & & $\nu_{14}$ & 910 & & $\nu_{14}$ & 802 & & $\nu_{14}$ & 716 & & $\nu_{14}$ & 588 & & $\nu_{14}$ & 658 \\
    $\nu_{15}$ & 796 & & $\nu_{15}$ & 985 & & $\nu_{15}$ & 878 & & $\nu_{15}$ & 790 & & $\nu_{15}$ & 634 & & $\nu_{15}$ & 658 \\
    $\nu_{16}$ & 876 & &            &     & & $\nu_{16}$ & 883 & & $\nu_{16}$ & 820 & & $\nu_{16}$ & 669 & & $\nu_{16}$ & 686 \\
    $\nu_{17}$ & 901 & &            &     & & $\nu_{17}$ & 989 & & $\nu_{17}$ & 827 & & $\nu_{17}$ & 676 & & $\nu_{17}$ & 729 \\
    $\nu_{18}$ & 976 & &            &     & & $\nu_{18}$ & 998 & & $\nu_{18}$ & 879 & & $\nu_{18}$ & 721 & & $\nu_{18}$ & 817 \\
    $\nu_{19}$ & 991 & &            &     & &            &     & & $\nu_{19}$ & 955 & & $\nu_{19}$ & 779 & & $\nu_{19}$ & 849 \\
           &     & &            &     & &            &     & & $\nu_{20}$ & 994 & & $\nu_{20}$ & 811 & & $\nu_{20}$ & 866 \\
           &     & &            &     & &            &     & &            &     & & $\nu_{21}$ & 909 & &          &     \\
           &     & &            &     & &            &     & &            &     & & $\nu_{22}$ & 916 & &          &     \\
           &     & &            &     & &            &     & &            &     & & $\nu_{23}$ & 962 & &          &     \\

    \hline
    \hline
    \end{tabular}
    \caption{Theoretical frequencies of vibrational modes for the $\widetilde X$ state of all molecules studied in this work. Only modes with frequencies smaller than 1000 \cm~ are listed. Values are given in units of \cm.}
    \label{table:freqall}
\end{table*}    
    
\begin{table*}
    \centering
    \setlength{\tabcolsep}{7pt}
    \renewcommand{\arraystretch}{1.5}
    \begin{tabular}{c c c c c c c c c}
    \hline
    \hline
    \multicolumn{4}{c}{CaOPh-34F} & &\multicolumn{4}{c}{CaOPh-345F}\\
    \cline{1-4}\cline{6-9} 
    modes & $\widetilde A \rightarrow \widetilde X$ & modes & $\widetilde B \rightarrow \widetilde X$& &modes& $\widetilde A \rightarrow \widetilde X$ & modes& $\widetilde B \rightarrow \widetilde X$     \\
    \cline{1-4}\cline{6-9}
    $0$ & 0.9674 & $0$ & 0.9831  && $0$ & 0.9755 & $0$ & 0.9886\\
    $\nu_5$ & 0.0136 & $\nu_5$ & 0.0073  & &$\nu_6$ & 0.0159 & $\nu_6$ & 0.0072\\
    $\nu_6$ & 0.0112 & $\nu_6$ & 0.0050 & &$\nu_8$ & 0.0031 & $\nu_{16}$ &0.0010\\
    $\nu_{11}$ & 0.0023 & $\nu_{11}$ & 0.0017  & &$\nu_{16}$ & 0.0018 & $\nu_{11}$ & 0.0009\\
    $\nu_{2}$ & 0.0012  & $\nu_{16}$ & 0.0009  & &$\nu_{24}$ & 0.0011 & $\nu_{8}$ & 0.0007\\
    $\nu_{16}$ & 0.0009 &$\nu_{10}$ & 0.0005  & &$\nu_{11}$ & 0.0009 & $\nu_{18}$ & 0.0005\\
    $\nu_{22}$ & 0.0007 &$\nu_{22}$ & 0.0003  & &$\nu_{27}$ & 0.0005 & $\nu_{24}$ & 0.0003\\
    $\nu_{26}$ & 0.0005 &$\nu_{26}$ & 0.0003  & &$\nu_{18}$ & 0.0004 & $\nu_{27}$ & 0.0002\\
    $\nu_{10}$ & 0.0004 &$\nu_{28}$ & 0.0002  & &$\nu_{29}$ & 0.0002 & 2$\nu_{2}$ & 0.0002\\
    $\nu_{15}$ & 0.0003 &2$\nu_{2}$ & 0.0002  & &2$\nu_{6}$ & 0.0002 & $\nu_{29}$ & 0.0001\\
    $\nu_{28}$ & 0.0003 &$\nu_{20}$ & 0.0001  & &$\nu_{21}$ & 0.0001 & $\nu_{21}$ & 0.0001\\
    $\nu_{20}$ & 0.0002 &$\nu_{15}$ & 0.0001  & &$\nu_{31}$ & 0.0001 & 2$\nu_{6}$ & 0.0001\\
    $\nu_{30}$ & 0.0002 &$\nu_{30}$ & 0.0001  & &2$\nu_{1}$ & 0.0001 & \\
    $\nu_5\nu_6$ & 0.0002 &$\nu_{27}$ & 0.0001  & &$\nu_{25}$ & 0.0001 \\
    $\nu_{24}$ & 0.0002 &2$\nu_{5}$ & 0.0001  & &$\nu_6\nu_8$ & 0.0001 \\
    $2\nu_{5}$ & 0.0001 & & & &&& \\
    $2\nu_{2}$ & 0.0001 & & & &&& \\
    $\nu_{27}$ & 0.0001 & & & &&& \\
    $2\nu_{6}$ & 0.0001 & & & &&& \\
    \hline
    \hline
    \end{tabular}
    \caption{Theoretical vibrational branching ratios of CaOPh-34F and CaOPh-345F above the level of $10^{-4}$. }
    \label{table:VBRs345f}
\end{table*}

\begin{table*}
    \centering
    \setlength{\tabcolsep}{7pt}
    \renewcommand{\arraystretch}{1.5}
    \begin{tabular}{c c c c c c c c c}
    \hline
    \hline
    \multicolumn{4}{c}{CaOPh-\textit{m}F} & &\multicolumn{4}{c}{CaOPh-\textit{m}CF$_3$}\\
    \cline{1-4}\cline{6-9} 
    modes & $\widetilde A \rightarrow \widetilde X$ & modes & $\widetilde B \rightarrow \widetilde X$& &modes& $\widetilde A \rightarrow \widetilde X$ & modes& $\widetilde B \rightarrow \widetilde X$     \\
    \cline{1-4}\cline{6-9}
    $0$ & 0.9676 & $0$ & 0.9824  && $0$ & 0.9699 & $0$ & 0.9828\\
    $\nu_5$ & 0.0222 & $\nu_5$ & 0.0122  & &$\nu_7$ & 0.0154 & $\nu_7$ & 0.0076\\
    $\nu_6$ & 0.0031 & $\nu_10$ & 0.0018 & &$\nu_9$ & 0.0031 & $\nu_{1}$ &0.0024\\
    $\nu_{10}$ & 0.0018 & $\nu_{6}$ & 0.0013  & &$\nu_{2}$ & 0.0028 & $\nu_{9}$ & 0.0015\\
    $\nu_{14}$ & 0.0010  & $\nu_{14}$ & 0.0007  & &$\nu_{10}$ & 0.0018 & $\nu_{15}$ & 0.0014\\
    $\nu_{1}$ & 0.0009 &$\nu_{25}$ & 0.0003  & &$\nu_{14}$ & 0.0013 & $\nu_{23}$ & 0.0008\\
    $\nu_{25}$ & 0.0008 &$\nu_{24}$ & 0.0002  & & $\nu_{8}$ & 0.0013 & $\nu_{10}$ & 0.0008\\
    $\nu_{24}$ & 0.0006 &$\nu_{27}$ & 0.0002  & &$\nu_{23}$ & 0.0009 & $\nu_{8}$ & 0.0005\\
    $2\nu_{24}$ & 0.0004 &$\nu_{18}$ & 0.0002  & &$\nu_{5}$ & 0.0006 & $\nu_{5}$ & 0.0004\\
    $\nu_{18}$ & 0.0003 &2$\nu_{5}$ & 0.0001  & &$\nu_{34}$ & 0.0005 & $\nu_{17}$ & 0.0003\\
    $2\nu_{1}$ & 0.0003 &2$\nu_{1}$ & 0.0001  & &$\nu_{19}$ & 0.0004 & $\nu_{2}$ & 0.0003\\
    $\nu_{29}$ & 0.0002 &$\nu_{19}$ & 0.0001  & &2$\nu_{2}$ & 0.0004 & $\nu_{19}$ & 0.0002\\
    $\nu_{22}$ & 0.0002 &$\nu_{29}$ & 0.0001  & &2$\nu_{7}$ & 0.0002 & $\nu_{34}$ & 0.0002\\
    $\nu_{27}$ & 0.0002 &$\nu_{26}$ & 0.0001  & &$\nu_{17}$ & 0.0002 &$\nu_{27}$ & 0.0002\\
    $\nu_{26}$ & 0.0001 &$\nu_{22}$ & 0.0001  & &$\nu_{36}$ & 0.0002 &$\nu_{36}$ & 0.0001\\
    $\nu_{21}$ & 0.0001 &$\nu_{22}$ & 0.0002 &&$\nu_{31}$ & 0.0001&$\nu_{35}$ & 0.0001\\
    $\nu_{4}\nu_5$ & 0.0001 &$\nu_{22}$ & 0.0002 &&$\nu_{38}$ & 0.0001&2$\nu_{2}$ & 0.0001 \\
    $2\nu_{2}$ & 0.0001 &$\nu_{22}$ & 0.0002 &&$\nu_{33}$ & 0.0001 &2$\nu_{7}$ & 0.0001\\
    $\nu_{19}$ & 0.0001 &$\nu_{22}$ & 0.0002 &&$\nu_{35}$ & 0.0001 &$\nu_{38}$ & 0.0001\\
    & & & & &$\nu_{32}$ & 0.0001 &2$\nu_{1}$ & 0.0001\\
    & & & & &$\nu_{14}$ & 0.0001 &&\\
    & & & & &$\nu_{26}$ & 0.0001 &&\\
    & & & & &$\nu_{1}$ & 0.0001 &&\\
    & & & & &$\nu_{12}$ & 0.0001 &&\\
    \hline
    \hline
    \end{tabular}
    \caption{Theoretical vibrational branching ratios of CaOPh-\textit{m}F and CaOPh-\textit{m}CF$_3$ above the level of $10^{-4}$. }
    \label{table:VBRsmF&mCF3}
\end{table*}
    
\begin{table*}
    \centering
    \setlength{\tabcolsep}{7pt}
    \renewcommand{\arraystretch}{1.5}
    \begin{tabular}{c c c c c c c c c}
    \hline
    \hline
    \multicolumn{4}{c}{CaOPh} & &\multicolumn{4}{c}{CaOPh-\textit{m}CH$_3$}\\
    \cline{1-4}\cline{6-9} 
    modes & $\widetilde A \rightarrow \widetilde X$ & modes & $\widetilde B \rightarrow \widetilde X$& &modes& $\widetilde A \rightarrow \widetilde X$ & modes& $\widetilde B \rightarrow \widetilde X$     \\
    \cline{1-4}\cline{6-9}
    $0$ & 0.9611 & $0$ & 0.9758  && $0$ & 0.9587 & $0$ & 0.9737\\
    $\nu_4$ & 0.0311 & $\nu_4$ & 0.0185  & &$\nu_6$ & 0.0175 & $\nu_6$ & 0.0110\\
    $\nu_9$ & 0.0027 & $\nu_9$ & 0.0027 & &$\nu_7$ & 0.0151 & $\nu_{7}$ &0.0090\\
    $\nu_{13}$ & 0.0018 & $\nu_{13}$ & 0.0012  & &$\nu_{12}$ & 0.0024 & $\nu_{12}$ & 0.0024\\
    $\nu_{23}$ & 0.0017  & $\nu_{23}$ & 0.0008  & &$\nu_{14}$ & 0.0016 & $\nu_{14}$ & 0.0014\\
    $2\nu_{4}$ & 0.0006 &$2\nu_{4}$ & 0.0003  & &$\nu_{27}$ & 0.0012 & $\nu_{27}$ & 0.0006\\
    $\nu_{28}$ & 0.0006 &$\nu_{26}$ & 0.0002  & &$\nu_{2}$ & 0.0006 & $\nu_{1}$ & 0.0003\\
    $\nu_{26}$ & 0.0006 &$\nu_{28}$ & 0.0001  & &$\nu_{18}$ & 0.0003 & $\nu_{33}$ & 0.0002\\
    $2\nu_{1}$ & 0.0002 &$2\nu_{2}$ & 0.0001  & &$2\nu_{2}$ & 0.0003 & $\nu_{21}$ & 0.0002\\
    $\nu_{4}\nu_{9}$ & 0.0001 &$\nu_{4}\nu_{9}$ & 0.0001  & &$\nu_{6}\nu_{7}$ & 0.0003 & $\nu_{18}$ & 0.0002\\
    $2\nu_{2}$ & 0.0001 && & &$\nu_{33}$ & 0.0002 & $\nu_{35}$ & 0.0001\\
    $\nu_{4}\nu_{23}$ & 0.0001 &&& &$\nu_{21}$ & 0.0002 & $\nu_{15}$ & 0.0001\\
    $\nu_{4}\nu_{13}$ & 0.0001 && & &$\nu_{25}$ & 0.0002 & $\nu_{30}$ & 0.0001\\
    & & & & &$\nu_{35}$ & 0.0002 &2$\nu_{2}$ & 0.0001\\
    & & & & &2$\nu_{6}$ & 0.0002 &$\nu_{6}\nu_{7}$ & 0.0001\\
    & & & & &$\nu_{28}$ & 0.0001 &$\nu_{25}$ & 0.0001\\
    & & & & &$\nu_{30}$ & 0.0001 &2$\nu_{6}$ & 0.0001 \\
    & & & & &$2\nu_{7}$ & 0.0001 &$\nu_{28}$ & 0.0001\\
    & & & & &$\nu_{15}$ & 0.0001 &$\nu_{34}$ & 0.0001\\
    & & & & &$\nu_{34}$ & 0.0001 &$\nu_{2}$ & 0.0001\\
    & & & & &$\nu_{29}$ & 0.0001 &$2\nu_{7}$ & 0.0001\\\
    & & & & &$2\nu_{3}$ & 0.0001 &&\\
    \hline
    \hline
    \end{tabular}
    \caption{Theoretical vibrational branching ratios of CaOPhand CaOPh-\textit{m}CH$_3$ above the level of $10^{-4}$. }
    \label{table:VBRscaoph&ch3}
\end{table*}

\begin{table*}
    \centering
    \setlength{\tabcolsep}{7pt}
    \renewcommand{\arraystretch}{1.5}
    \begin{tabular}{c c c c c c c c}
    \hline
    \hline
    \multicolumn{2}{l}{decays from A (v=0)} &  & \multicolumn{2}{l}{decays from A (v=1) of mode $\nu_1$} &  & \multicolumn{2}{l}{decays from A (v=1) of mode $\nu_4$} \\
    \cline{1-2}\cline{4-5}\cline{7-8}
    modes & VBR    &  & modes   & VBR    &  & modes    & VBR    \\
    \hline
    0     & 0.9611 &  & $\nu_1$     & 0.9609 &  & $\nu_4$      & 0.8963 \\
    $\nu_4$    & 0.0311 &  & $\nu_1\nu_4$    & 0.0311 &  & $2\nu_4$      & 0.0598 \\
    $\nu_9$    & 0.0027 &  & $\nu_1\nu_9$    & 0.0027 &  & 0        & 0.0354 \\
    $\nu_{23}$   & 0.0017 &  & $\nu_1\nu_{23}$   & 0.0017 &  & $\nu_4\nu_9$  & 0.0025 \\
    $\nu_{13}$   & 0.0018 &  & $\nu_1\nu_{13}$   & 0.0018 &  & $\nu_4\nu_{23}$ & 0.0015 \\
    2$\nu_4$   & 0.0006 &  & $\nu_12\nu_4$  & 0.0006 &  & $4\nu_4$      & 0.0017 \\
    $\nu_{28}$   & 0.0003 &  & $3\nu_1$     & 0.0005 &  & $\nu_4\nu_{13}$ & 0.0016 \\
    $\nu_{26}$   & 0.0003 &  & $\nu_1\nu_{28}$   & 0.0003 &  & $\nu_4\nu_{28}$ & 0.0003 \\
    2$\nu_1$   & 0.0002 &  & $\nu_1\nu_{26}$   & 0.0003 &  & $\nu_4\nu_{26}$ & 0.0003 \\
    $\nu_4\nu_9$  & 0.0001 &  & $\nu_1\nu_4\nu_9$  & 0.0001 &  & $2\nu_4\nu_{9}$  & 0.0002 \\
    2$\nu_2$   & 0.0001 &  & $\nu_12\nu_2$  & 0.0001 &  & $2\nu_1\nu_{4}$  & 0.0002 \\
    $\nu_4\nu_{23}$ & 0.0001 &  & $\nu_1\nu_4\nu_{23}$ & 0.0001 &  & $2\nu_4\nu_{23}$ & 0.0001  \\  
    \hline
    \hline
    \end{tabular}
    \caption{Theoretical vibrational branching ratios from different vibrational levels of $\widetilde{A}$ state of CaOPh above the level of $10^{-4}$. }
    \label{table:diagonalVBR}
\end{table*}

\renewcommand*{\thefigure}{S\arabic{figure}}
\setcounter{figure}{0}    
\begin{figure*}
    \centering
    \includegraphics[width = 0.95\textwidth]{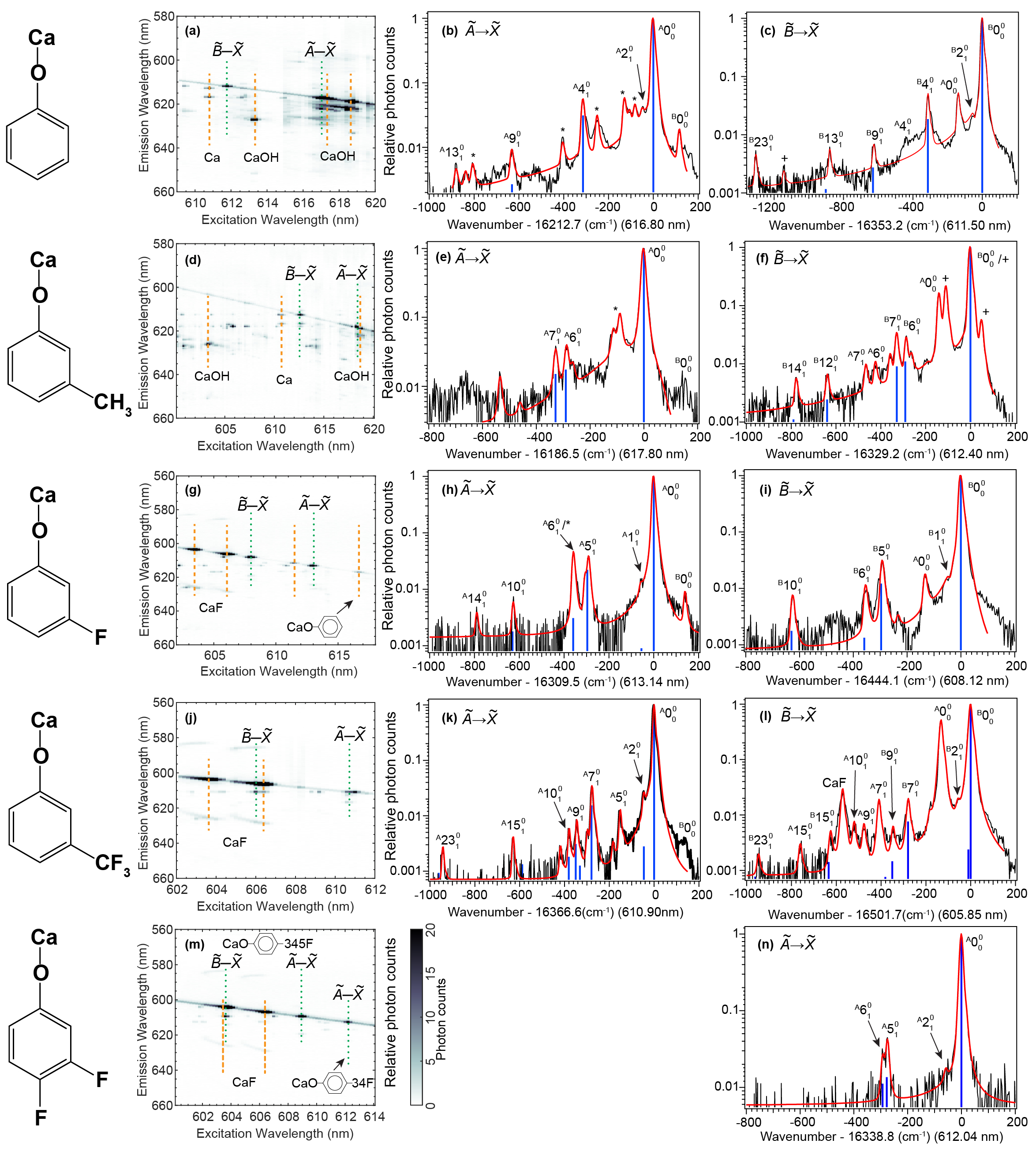}
    \caption{2D spectra and dispersed LIF spectra of all species. In the 2D spectra, the orange dashed lines mark features due to CaOH or CaF, while the green dotted lines indicate features from CaOPh-X species. In the corresponding dispersed LIF spectra, the experimental curves (black) are fitted with Pearson functions (red). The blue sticks illustrate the vibrational branching ratios of different vibrational modes. The assignments of resolved vibrational peaks are also given. The symbols * and + indicate features due to CaOH and Ca, respectively.}
    \label{fig.caoph-all-dlif}
\end{figure*}

\begin{figure*}
    \centering
    \includegraphics[scale=0.4]{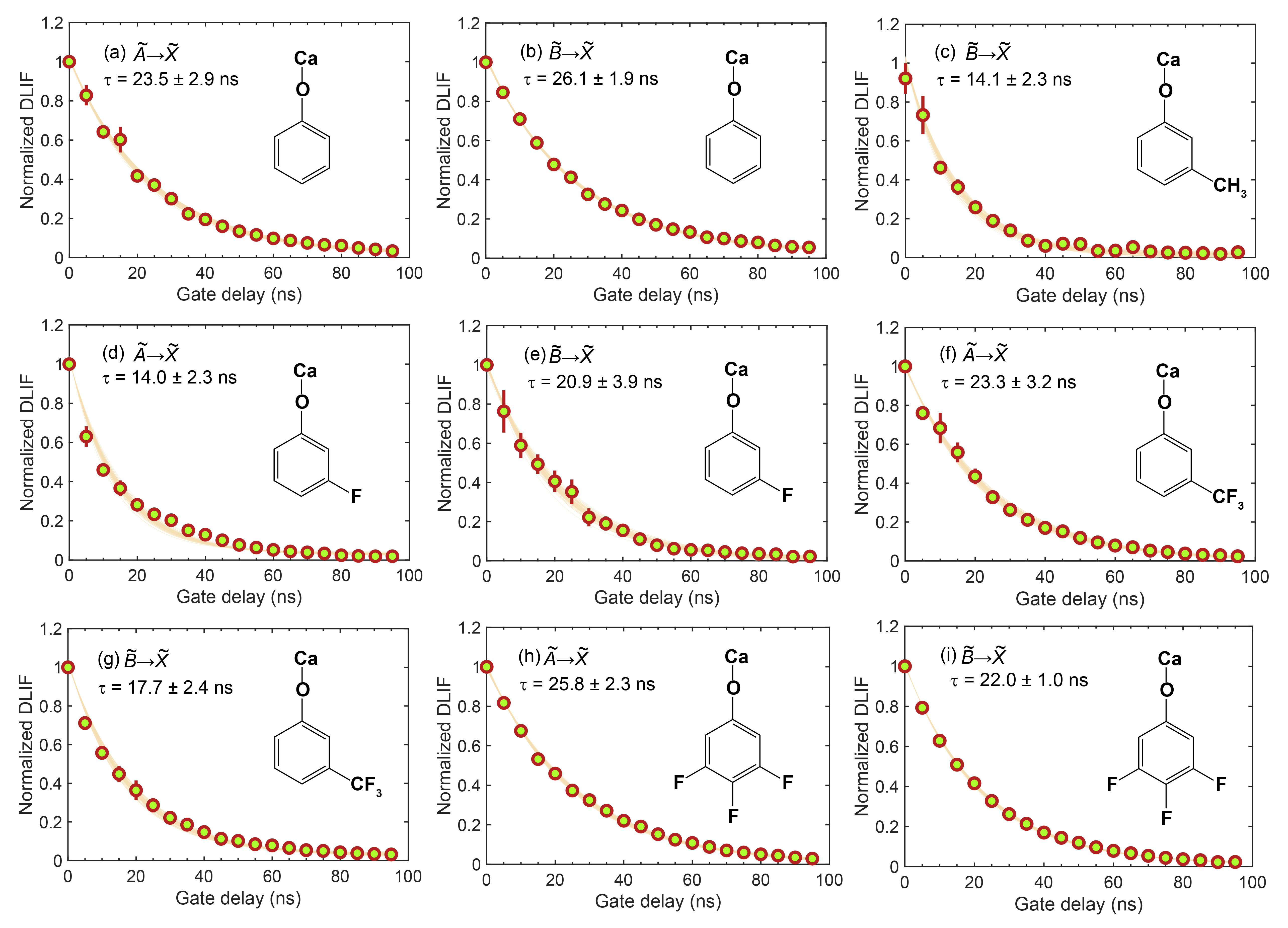}
    \caption{Fluorescence decay traces used to determine radiative lifetimes for all species observed in this work. The decay trace of each transition has been measured 2-4 times. The experimental data points show the averaged values of all normalized measurements at the same gate delay and the error bar of each point represents the standard deviations. The experimental points are fitted with an exponential function to extract the lifetime and the corresponding error bars are obtained by bootstrapping the data. 
    }
    \label{fig:lifetime}
\end{figure*}

\begin{figure*}
    \centering
    \includegraphics[scale=1]{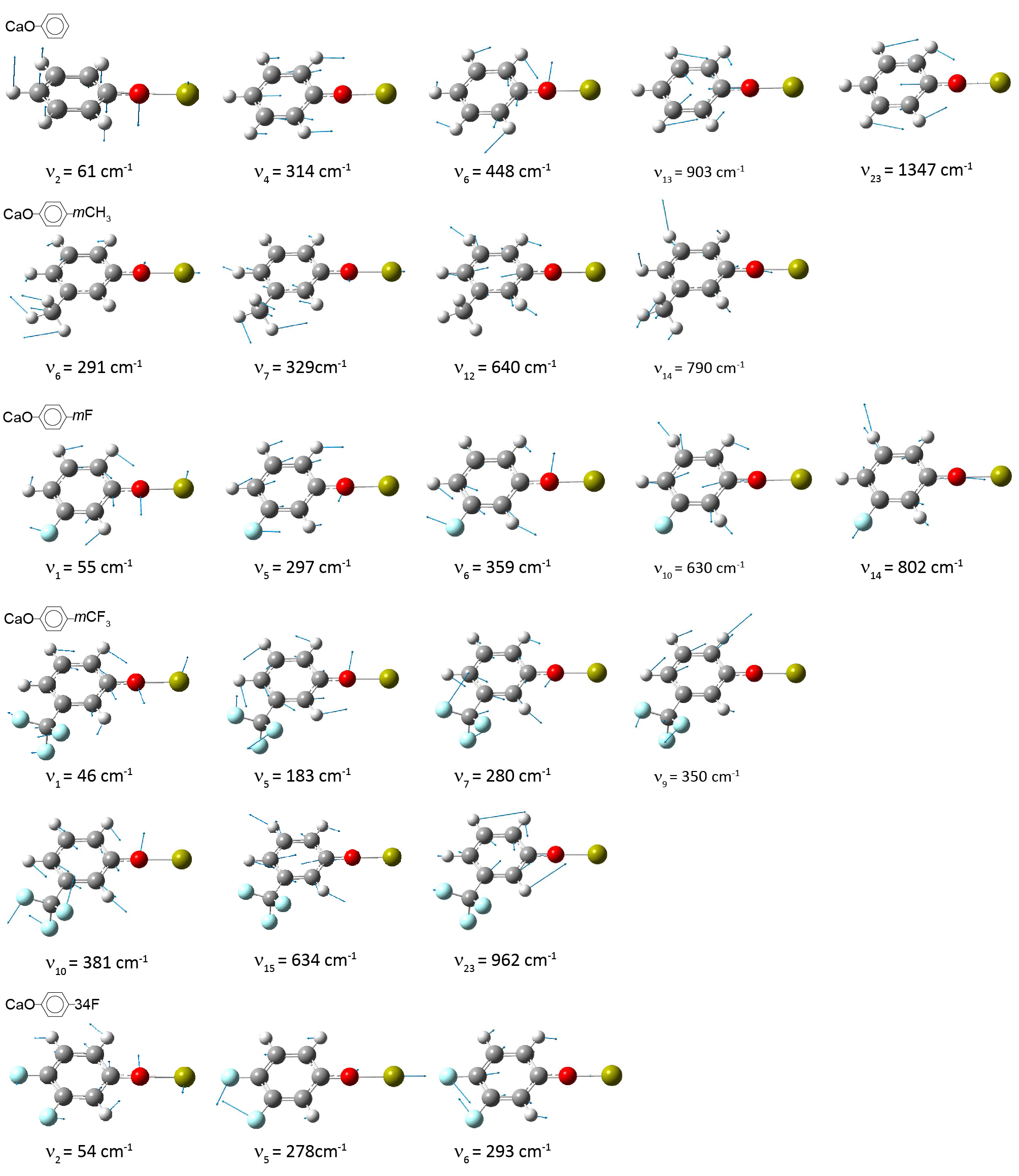}
    \caption{Schematic illustrations of resolved fundamental normal vibrational modes. The arrows indicate the direction of vibrational displacements. The corresponding theoretical frequencies are also given.
    }
    \label{fig:vib.modes}
\end{figure*}

\end{document}